\documentclass[a4paper,11pt]{article}
\pdfoutput=1

\usepackage{jcappub}

\usepackage[T1]{fontenc}

\usepackage[english]{babel}
\usepackage{amsfonts}
\usepackage{txfonts}
\usepackage{graphicx}
\usepackage{amssymb}
\usepackage{ae,aecompl}
\usepackage{fancyhdr}
\usepackage{multicol}
\usepackage{layout}
\usepackage{graphicx}
\usepackage{multirow}
\usepackage{times}
\usepackage{textcomp}
\usepackage{cprotect}

\usepackage{soul}

\usepackage[outdir=./figures/]{epstopdf}
\usepackage[export]{adjustbox}
    
\graphicspath{{./figures/}}

\newif\ifAMStwofonts
\AMStwofontstrue

\title{Halo collapse: virialization by shear and rotation in dynamical dark-energy models. Effects on weak-lensing 
peaks}

\author[a,1]{Francesco Pace,\note{Corresponding author.}}
\author[b]{Carlo Schimd,}
\author[c]{David F. Mota}
\author[d,e,f]{and Antonino Del Popolo}

\affiliation[a]{Jodrell Bank Centre for Astrophysics, School of Natural Sciences, Department of Physics and Astronomy, 
The University of Manchester, Manchester, M13 9PL, United Kingdom}
\affiliation[b]{Aix  Marseille  Univ,  CNRS, CNES, LAM,  Marseille,  France}
\affiliation[c]{Institute of Theoretical Astrophysics, University of Oslo, P.O Box 1029 Blindern, N-0315 Oslo, Norway}
\affiliation[d]{Dipartimento di Fisica e Astronomia, University Of Catania, Viale Andrea Doria 6, 95125, Catania, 
Italy}
\affiliation[e]{Institute of Astronomy of Russian Academy of Sciences, 48 Pyatnitskaya str., Moscow 119017, Russia}
\affiliation[f]{INFN sezione di Catania, Via S. Sofia 64, I-95123 Catania, Italy}

\emailAdd{francesco.pace@manchester.ac.uk}
\emailAdd{carlo.schimd@lam.fr}
\emailAdd{d.f.mota@astro.uio.no}
\emailAdd{adelpopolo@oact.inaf.it}

\abstract{
The evolution of the virial overdensity $\Delta_{\rm vir}$ for $\Lambda$CDM and seven dynamical dark-energy models 
is investigated in the extended spherical collapse model (SCM). Here the virialization process is naturally achieved 
by introducing shear and rotation instead of using the virial theorem. We generalise two approaches proposed in the 
literature and show that, regardless of the dark-energy model, the new virialization term can be calibrated on the 
peculiar velocity of the shell as measured from Einstein-de Sitter simulations. The two virialization recipes 
qualitatively reproduce the features of the ordinary SCM, i.e., a constant $\Delta_{\rm vir}$ for the EdS model and 
time-variation for dark-energy models, but without any mass dependence. Depending on the actual description of 
virialization and on the dark-energy model, the value of $\Delta_{\rm vir}$ varies between 10 and 40 percent. 
We use the new recipes to predict the surface-mass-density profile of dark matter haloes and the number of convergence 
density peaks for \textit{LSST}- and \textit{Euclid}-like weak lensing surveys.
}


\arxivnumber{1811.12105}

\begin{document}

\label{firstpage}

\maketitle
\flushbottom

\section{Introduction}\label{sect:intro}
A major drawback of the spherical collapse model (SCM) \citep{Tomita1969,Gunn1972} is the singularity attained in a 
finite time at collapse, consequence of the spherical symmetry of the perturbation. Deviations from spherical symmetry 
and relaxation processes such as phase mixing, chaotic mixing, violent relaxation and Landau damping lead to 
virialized non-singular structures, in which the kinetic energy is converted into random motions of particles 
resulting in a finite radius of the final overdensity. Instead, in the SCM a virialization condition is forced 
\textit{a posteriori} by means of the virial theorem, which for the Einstein-de Sitter (EdS) model yields the 
well-known time-independent exact value $\Delta_{\rm vir}\simeq 178$ of the virial overdensity corresponding to the 
linearly-extrapolated overdensity $\delta_{\rm c}\simeq 1.686$ at collapse. 
Structures evolve until this value is reached, which corresponds to a finite value of the virial radius. 
For dark-energy cosmologies, including the $\Lambda$CDM model, the modelling is more complex and a full analytical 
solution has not been established yet. Energy conservation and virialization in presence of dark-energy are still 
unclear, and are expected to depend on the degree of smoothness of the dark-energy fluid 
\citep{Wang1998,Iliev2001,Battye2003,Horellou2005,Maor2005,Wang2006,Batista2017}.

An interesting alternative and physically motivated approach is the extended SCM (hereafter ESCM) proposed by 
\cite{Engineer2000} and refined by \cite{Shaw2008}. They obtained a natural virialization by means of additional 
shear and rotation terms written as a Taylor expansion in $1/\delta$, which encode the correct dynamics for large 
values of the density contrast $\delta$. The new terms finally lead to aspherical structures better fitting both 
observations and $N$-body simulations, which actually show triaxial dark matter haloes and galaxy clusters 
\citep{Limousin2013}. 
Alternatively, the additional terms can be modelled as proportional to $\delta$ 
\citep{DelPopolo2013a,DelPopolo2013b,Pace2014b}, resulting in a slower mass-dependent collapse as in the ellipsoidal 
model \citep{Bond1996a,Bond1996b,Ohta2003,Ohta2004,Angrick2010}, with higher values of $\Delta_{\rm vir}$ and 
$\delta_{\rm c}$ with respect to the standard SCM.

In this work, we compare the standard approach in which virialization is not native, as in the SCM formalism, with two 
physically motivated recipes for virialization introduced by \cite{Engineer2000} and \cite{Shaw2008}, explicitly 
worked out for seven dark-energy models alternative to the $\Lambda$CDM. Assuming dark matter haloes with a 
Navarro-Frenk-White (NFW) profile \cite{Navarro1997}, the impact of the new virialization recipes is assessed by 
calculating the radial surface-mass-density as measured by weak-lensing and estimating the number of convergence 
density peaks for \textit{LSST}-\footnote{https://www.lsst.org/} and 
\textit{Euclid}-\footnote{https://www.euclid-ec.org} like weak lensing surveys. 
The plan of this work is as follows: In Section~\ref{sect:SCM} we summarise the equations of the ESCM focusing on 
cosmologies with dark-energy and we detail the shear-rotation-induced virialization mechanism and generalise the 
methods by \cite{Engineer2000,Shaw2008} to the $\Lambda$CDM and to arbitrary smooth dark-energy models. In 
Section~\ref{sect:analysis} we compare the different approaches by showing the evolution of the virial overdensity 
$\Delta_{\rm vir}$ as a function of time for different dark-energy models including the $\Lambda$CDM scenario. 
Section~\ref{sect:lensing} is dedicated to the weak-lensing observables. Section~\ref{sect:conclusions} summarises our 
findings.

The formalism of the ordinary SCM in an EdS universe is reminded in Appendix~\ref{appendix:SCM}, while the numerical 
implementation of the ESCM with dark-energy models is outlined in Appendix~\ref{sect:implementation}. Throughout the 
paper a spatially flat universe is considered, with matter density parameter $\Omega_{\rm m,0}=0.3$.


\section{The extended spherical collapse model}\label{sect:SCM}

\subsection{Top-hat dynamics with shear and rotation}

The equations of motion for the density contrast $\delta\equiv\delta\rho_{\rm m}/\bar{\rho}_{\rm m}>0$, with 
$\bar{\rho}_{\rm m}$ the matter background density, can be derived following the fluid approach.\footnote{Since we 
consider the formation of structures at late times, we neglect radiation and consider matter ($\bar{\rho}_{\rm m}$) as 
made of baryons ($\bar{\rho}_{\rm b}$) and cold dark matter ($\bar{\rho}_{\rm cdm}$).} 
Combining the continuity, Euler and Poisson equations for a collisionless fluid under the top-hat approximation, 
$\delta$ solves the non-linear differential equation \citep{Ohta2003,Pace2010,Pace2017a}
\begin{equation}\label{eqn:nldelta}
 \ddot{\delta} + 2H\dot{\delta} - \frac{4}{3}\frac{\dot{\delta}^2}{1+\delta} - 
 \frac{3}{2}H^2\Omega_{\rm m}(t)\delta(1+\delta) = (1+\delta)(\sigma^2-\omega^2)\,,
\end{equation}
where a dot represents the derivative with respect to the cosmic time, $H$ and $\Omega_{\rm m}(t)$ are the Hubble and 
matter density parameters, and $\sigma^2\equiv\sigma_{ij}\sigma^{ij}$ and $\omega^2\equiv\omega_{ij}\omega^{ij}$ the 
squared amplitude of the shear and rotation tensors, respectively. The shear $\sigma_{ij}$ and rotation $\omega_{ij}$ 
tensors represent the symmetric traceless and anti-symmetric component of the derivative of the peculiar velocity 
$\boldsymbol{u}$,
\begin{equation}\label{eqn:veltensor}
 \partial_i u_j = \frac{1}{3}\theta\delta_{ij}+\sigma_{ij}+\omega_{ij}\,,
\end{equation}
where the trace $\theta\equiv\vec{\nabla}\cdot\boldsymbol{u}$ accounts for the isotropic expansion and the other two 
terms are defined by
\begin{equation}
 \sigma_{ij} = \frac{1}{2}\left(\partial_i u_j + \partial_j u_i\right) - \frac{1}{3}\theta\delta_{ij}\,, \quad 
 \omega_{ij} = \frac{1}{2}\left(\partial_i u_j - \partial_j u_i\right)\,.
\end{equation}
From equation~(\ref{eqn:veltensor}), one obtains 
$\vec{\nabla}\cdot[(\boldsymbol{u}\cdot\vec{\nabla})\boldsymbol{u}] = \frac{1}{3}\theta^2 + \sigma^2 - \omega^2$, 
which enters into equation~(\ref{eqn:nldelta}) via the Euler equation.\footnote{Sometimes in the literature the 
rotation vector $\Omega^k$ is used, related to the rotation tensor by the relation 
$\omega_{ij}=\epsilon_{ijk}\Omega^k$ with $\epsilon_{ijk}$ the totally anti-symmetric Levi-Civita tensor. Accordingly 
$\omega^2=2\Omega^2$, paying attention to avoid confusion with the density parameter.}

\subsection{Virialization by parametric models (approach I)}\label{sect:SCM:approach1}

The approach outlined by \cite{Engineer2000} and further developed by \cite{Shaw2008} is here applied to investigate 
the extended SCM in presence of smooth dark-energy component. The validity of this extension depends on the difference 
between the two-point correlation functions of the models \citep[see][for a more detailed discussion]{Engineer2000}.

Equation~(\ref{eqn:nldelta}) can be written in terms of the mass and radius of the (spherical) perturbation, noting 
that
\begin{equation}\label{eqn:delta_R}
 1 + \delta = \frac{2GM_{\rm m}a^3}{\Omega_{\rm m0}H_0^2R^3} \equiv \lambda\frac{a^3}{R^3}\,,
\end{equation}
with $M_{\rm m}$ the (dark matter) mass content, $G$ the Newtonian gravitational constant, $a$ the scale factor, and 
subscript 0 denoting quantities evaluated today. Dealing with aspherical perturbations because of shear and rotation, 
$R$ must be interpreted as an effective length scale. One obtains
\begin{equation}\label{eqn:Rddot}
 \ddot{R}  = -\frac{GM_{\rm m}}{R^2} - \frac{GM_{\rm de}}{R^2}(1+3w_{\rm de}) - \frac{R}{3}H^2S \,,
\end{equation}
where $S=\tilde{\sigma}^2-\tilde{\omega}^2\equiv H^{-2}(\sigma^2-\omega^2)$ is the dimensionless function that encodes 
shear and rotation \citep{Engineer2000,Shaw2008} and $M_{\rm de}=\frac{4\pi}{3}\bar{\rho}_{\rm de}R^3$ is the mass of 
the dark-energy component enclosed in the spheroid, $\bar{\rho}_{\rm de}$ and $w_{\rm de}$ being respectively its 
background density and equation-of-state (not necessarily constant). This term is obtained from the $\dot{H}/H^2$ term 
and is important when discussing the correction of dark-energy perturbations to the halo mass function 
\citep{Batista2013}. When virialization occurs, $R$ tends to the constant value $R_\mathrm{vir}$ and $\dot{R}\to0$, 
therefore setting $\ddot{R}\approx 0$, we find
\begin{equation}\label{eqn:S1}
 S \approx -\frac{3GM_{\rm m}}{H^2R^3}\left[1+\frac{M_{\rm de}}{M_{\rm m}}(1+3w_{\rm de})\right] \approx - 
            \frac{3}{2}\Omega_{\rm m}(a)(1+\delta)\,.
\end{equation}
The first expression generalises the result in \cite{Engineer2000} and the second approximation is valid when the dark 
energy is sub-dominant with respect to the dark matter component.

Using instead the $e$-fold time $\ln{a}$ as time variable, which is more appropriate when dealing with dark-energy 
models and useful for the numerical implementation \citep{Pace2017a}, Equation~(\ref{eqn:nldelta}) reads
\begin{equation}\label{eqn:nldeltaa}
   \delta^{\prime\prime} + \left(2+\frac{H^{\prime}}{H}\right)\delta^{\prime} - 
   \frac{4}{3}\frac{{\delta^{\prime}}^2}{1+\delta} - 
   \frac{3}{2}\Omega_{\rm m}(a)\delta(1+\delta) = (1+\delta)S \,,
\end{equation}
where the prime indicates the derivative with respect to $\ln{a}$. When virialization takes place, $\delta$ tends to a 
constant value and $\delta^\prime\approx0$ so that the previous equation yields
\begin{equation}\label{eqn:S2}
 S \approx -\frac{3}{2}\Omega_{\rm m}(a)\delta\,.
\end{equation}
The expressions (\ref{eqn:S1}) and (\ref{eqn:S2}) coincide in the limit $\delta\gg1$, but for intermediate values they 
lead to different results. Therefore hereafter we will show results for $\Delta_{\rm vir}$ using both expressions, 
referred respectively as model S$_1$ and S$_2$.

The full expressions for the virial term $S\equiv S(\delta)$ accounting also for the Taylor expansion in $1/\delta$ up 
to the second order are \cite{Engineer2000}
\begin{subequations}\label{eqn:S12}
 \begin{eqnarray}
  S_{1} & = & -\frac{3}{2}\Omega_{\rm m}(a)(1+\delta) - \frac{A}{\delta} + \frac{B}{\delta^2}\,, \label{eqn:S1a}\\
  S_{2} & = & -\frac{3}{2}\Omega_{\rm m}(a)\delta - \frac{A}{\delta} + \frac{B}{\delta^2}\,, \label{eqn:S2a}
\end{eqnarray}
\end{subequations}
where $A$ and $B$ are numerical constants. The evolution equation~(\ref{eqn:nldeltaa}) then reads
\begin{subequations}\label{eqn:full}
 \begin{eqnarray}
  \delta^{\prime\prime} & + & \left(2+\frac{H^{\prime}}{H}\right)\delta^{\prime} - 
      \frac{4}{3}\frac{{\delta^{\prime}}^2}{1+\delta} + 
      \frac{3}{2}\Omega_{\rm m}(a)(1+\delta) = (1+\delta)\left(-\frac{A}{\delta}+\frac{B}{\delta^2}\right)\,,\\
  \delta^{\prime\prime} & + & \left(2+\frac{H^{\prime}}{H}\right)\delta^{\prime} - 
  \frac{4}{3}\frac{{\delta^{\prime}}^2}{1+\delta} = (1+\delta)\left(-\frac{A}{\delta}+\frac{B}{\delta^2}\right)\,.
 \end{eqnarray}
\end{subequations}

Following \cite{Engineer2000}, in the limit $\delta\gg 1$ and rescaling $\delta\mapsto A\delta$ the r.h.s. term reads 
$-1+q/\delta$ with $q=B/A^2$, proving that the model only depends on the parameter $q$.\footnote{Note that we corrected 
the typo in \cite{Engineer2000} where it is erroneously written $\delta\mapsto\delta/A$ and the variable $b$ in 
Equation~(27) which should read as $a$.}
The equations fixing the overdensity at virialization will then read
\begin{subequations}\label{eqn:deltaS}
 \begin{eqnarray}
  \delta^{\prime\prime} & + & \left(2+\frac{H^{\prime}}{H}\right)\delta^{\prime} - 
  \frac{4}{3}\frac{{\delta^{\prime}}^2}{1+\delta} + 
  \frac{3}{2}\Omega_{\rm m}(a)(1+\delta) = (1+\delta)\left(-\frac{1}{\delta}+\frac{q}{\delta^2}\right)\,,\\
  \delta^{\prime\prime} & + & \left(2+\frac{H^{\prime}}{H}\right)\delta^{\prime} - 
  \frac{4}{3}\frac{{\delta^{\prime}}^2}{1+\delta} = (1+\delta)\left(-\frac{1}{\delta}+\frac{q}{\delta^2}\right)\,.
 \end{eqnarray}
\end{subequations}
Note that, albeit the two virialization expressions have the same limit, the exact dynamics of the models is slightly 
different, as shown in the following. In particular, the turn-around epoch and the maximum value of the effective 
radius will change in the two prescriptions. The reason is not only a consequence of the particular form of the 
equations of motion, but of the assumptions behind these models: model S$_1$ is derived assuming a constant final 
radius, while model S$_2$ a constant final density. While both conditions must be true when the object virialises (and 
in fact S$_1$=S$_2$ for $\delta\gg 1$), the intermediate details will differ and this is captured by the exact form of 
the equations. We finally notice that the equation of motion with recipe S$_2$ is the same expression studied by 
\cite{Engineer2000}.

Still in the limit $\delta\gg 1$, these equations written for $R$ read
\begin{subequations}
 \begin{eqnarray}
  R^{\prime\prime}+\frac{H^{\prime}}{H}R^{\prime} & = & -\frac{1}{2}\Omega_{\rm de}(a)(1+3w_{\rm de})R + 
                   \frac{R}{3}\left[A\frac{\Omega_{\rm m,0}}{2a^3}\frac{R^3}{r^3}
                   -qA^2\frac{\Omega_{\rm m,0}^2}{4a^6}\frac{R^6}{r^6}\right]\\
  R^{\prime\prime}+\frac{H^{\prime}}{H}R^{\prime} & = & -\frac{1}{2}\left[\Omega_{\rm m}(a)
                                                        +\Omega_{\rm de}(a)(1+3w_{\rm de})\right]R + 
                   \frac{R}{3}\left[A\frac{\Omega_{\rm m,0}}{2a^3}\frac{R^3}{r^3}
                  -qA^2\frac{\Omega_{\rm m,0}^2}{4a^6}\frac{R^6}{r^6}\right],
 \end{eqnarray}
\end{subequations}
where the quantity in brackets is the virialization term and we set $r^3=GM_{\rm m}/H_0^2$. 
By defining $R=r_{\rm vir}y$, the only undefined constants are $A$ and the virialization radius $r_{\rm vir}$. 
Since $A$ is arbitrary, it can be set to $A=r^3/r_{\rm vir}^3=GM_{\rm m}/H_0^2r_{\rm vir}^3$ so that the previous 
equations finally become
\begin{subequations}\label{eqn:yq}
 \begin{eqnarray}
  y^{\prime\prime}+\frac{H^{\prime}}{H}y^{\prime} & = & -\frac{1}{2}\Omega_{\rm de}(a)(1+3w_{\rm de})y + 
                                                         \frac{y}{3}\left[\frac{\Omega_{\rm m,0}}{2a^3}y^3
                                                        -q\frac{\Omega_{\rm m,0}^2}{4a^6}y^6\right]\,,\\
  y^{\prime\prime}+\frac{H^{\prime}}{H}y^{\prime} & = & 
                    -\frac{1}{2}\left[\Omega_{\rm m}(a)+\Omega_{\rm de}(a)(1+3w_{\rm de})\right]y + 
                     \frac{y}{3}\left[\frac{\Omega_{\rm m,0}}{2a^3}y^3-q\frac{\Omega_{\rm m,0}^2}{4a^6}y^6\right]\,.
 \end{eqnarray}
\end{subequations}
These equations agree with Eq.~(30) of \cite{Engineer2000}.

\subsection{Virialization fitting N-body dynamics (approach II)}
\label{sect:SCM:approach2}

The former approach is not valid for small overdensity because of the divergence of the terms proportional to 
$1/\delta$ and $(1/\delta)^2$. As suggested by \cite{Shaw2008}, one can instead consider the solution of the EdS model 
and replace the term $(\eta-\sin{\eta})$ that appears in the temporal part by a generic function $T(\eta)$ since only 
the time needed for collapse does change. Denoting by $T_\tau$ and $T_{\tau\tau}$ the first and second derivative of 
$T$ with respect to $\tau\equiv\tau(\eta)=\eta-\sin{\eta}$, one then obtains
\begin{equation}
 \ddot{R} = -\frac{GM}{R^2}\left[\frac{1}{{T_\tau}^2}+\sin{\eta}(1-\cos{\eta})
                      \frac{T_{\tau\tau}}{{T_\tau}^3}\right]\,.
\end{equation}
By comparison with Equation~(\ref{eqn:Rddot}), one then obtains the expression for the deviations from spherical 
symmetry (dubbed as model S$_{\rm T}$)
\begin{equation}\label{eqn:ST}
 S(\delta) = \frac{3}{2}(1+\delta)\left[\frac{1}{{T_\tau}^2}+\sin{\eta}(1-\cos{\eta})
             \frac{T_{\tau\tau}}{{T_\tau}^3}-1\right]\,,
\end{equation}
which requires a (parametric) relation between $T(\eta)$ and $\delta$ to be fully defined. In generic cosmological 
scenarios where dark-energy dominates at late times, this is
\begin{equation}\label{eqn:delta_T}
 1+\delta = \frac{9}{2}\tilde{f}(a)\frac{T^2(\eta)}{[1-\cos{(\eta)}]^3}\,,
\end{equation}
in which $\tilde{f}(a)=4/[9 t^2\Omega_{\rm m}(a)H^2(a)]$. A functional form of $T(\eta)$ was obtained by 
\cite{Shaw2008} for an EdS cosmology by fitting the $N$-body simulation of \cite{Hamilton1991}, which yielded
\begin{equation}\label{eqn:Ttau}
 T[\eta(\tau)] \equiv T(\tau) = \tau + \frac{3.468(\tau_f-\tau)^{-1/2}\,\mathrm{e}^{-15(\tau_f-\tau)/\tau}}
                             {1+0.8(\tau_f-\tau)^{1/2}-0.4(\tau_f-\tau)}\,,
\end{equation}
with $\tau_f=5.516$. As proved by \cite{Engineer2000}, this expressions is extremely good for $\delta\gtrsim 15$ and 
can be safely used to describe the evolution of the perturbation prior to the turn-around also in cosmologies with 
smooth dark-energy.\footnote{Note that the best fit value for the virial radius is now $R_{\rm vir}/R_{\rm ta}=0.5896$, 
in agreement with \cite{Hamilton1991} who found $R_{\rm vir}/R_{\rm ta}\approx 0.56$ while \cite{Engineer2000} found 
$R_{\rm vir}/R_{\rm ta}\approx 0.65$.}

As shown by \cite{Smith2003}, stable clustering is not very realistic. Nevertheless, as more carefully discussed in 
the next subsection, this does not represent a fundamental problem for this work, as the fitting expression 
in Equation~(\ref{eqn:Ttau}) is a direct fit to EdS simulations and does do not relay on the stable clustering 
assumption, which was instead used by \cite{Engineer2000} as guidance to interpret the results.

A final comment is worth to be mentioned. The works by \cite{Engineer2000} and \cite{Shaw2008} achieved the 
virialization with $\sigma^2-\omega^2\propto 1/\delta$. 
Instead, with $\sigma^2-\omega^2\propto\delta$ though, including virialization a posteriori, 
\cite{DelPopolo2013a,DelPopolo2013b,Pace2014b} obtained a delayed collapse but a virial overdensity substantially 
increased, while using the Zel'dovich approximation \cite{Reischke2016a,Pace2017,Reischke2018} found a negligible 
effect on $\Delta_{\rm vir}$ and a one percent variation for $\delta_{\rm c}$.

\subsection{Comment on stable clustering}
Before investigating in detail the properties of the models subject of this study, we deem important to discuss more 
one of the assumptions of the models proposed by \cite{Engineer2000}. In this work, the main hypothesis is that the 
virialization term is only a function of the matter overdensity, $S=S(\delta)$, and this idea has been used later to 
study structure formation using the formalism of the spherical collapse model augmented by a shear and rotation term 
\citep{DelPopolo2013a,DelPopolo2013b,Pace2014b,Reischke2016a,Pace2017,Reischke2018}, the main difference being that 
$S\propto\delta$ and not constructed with the idea of virialize the collapse; see Sect.~\ref{sect:SCM:approach2}. In 
\cite{Engineer2000}, this assumption is justified by considering the stable-clustering Ansatz in the highly non-linear 
regime. 
According to stable clustering, the rescaled pairwise velocity $h$ depends on time and (comoving) scale via the 
overdensity $\delta$ since the halo density profile is approximated near the high-peaks with the two-point correlation 
function $\bar{\xi}$, i.e., $\delta=\bar{\xi}$. 
Therefore a fundamental question to answer is whether stable clustering is still a viable assumption.

Firstly note that the relationship linking $\bar{\xi}$ to $h$ is of statistical nature (as it is for the approaches of 
\cite{Engineer2000} and \cite{Shaw2008} as noticed by \cite{Rasanen2008}) and only holds in the non-linear regime where 
$\delta>1$, which is the one we always consider in this work. In fact, for $\delta\lesssim 15$ we use the standard 
equations of the spherical collapse model. The authors of \citep{Engineer2000} also noticed that 
$\bar{\xi}\propto\delta$ using Lagrangian coordinates and this is always satisfied, unless 
$\delta\lesssim 1$.

The apparently weak point is that these conclusions are drawn based on the use of Einstein-de Sitter cosmological 
simulations. In reality, they also hold for non-minimally coupled dark energy models (while it is not necessarily the 
case for modified gravity models). This happens for a series of reasons: if dark energy is minimally coupled to gravity 
and no screening mechanism is involved (see the discussion in Sect.~\ref{sect:DeltaV}), structure formation is still 
very well approximated by the Newtonian dynamics and therefore there are no major differences from a $\Lambda$CDM 
evolution (except for timing). At the same time, being this the case, dark energy is the dominant component only at 
very late times and becomes important for $z\lesssim 1$, hence for most of the cosmic history the evolution is 
approximately as that of an Einstein-de Sitter model (this is also clearly seen by the evolution of 
$\Delta_{\rm vir}$) in Fig.~(\ref{fig:DeltaV}). We notice though that the assumption of smooth dark energy is not 
crucial, as it was shown in \cite{Batista2013} that even for early dark energy models, when perturbations are 
considered, the dynamics is more similar to $\Lambda$CDM. Accordingly, in a future work we plan to extend the approach 
outlined by \cite{Engineer2000} to clustering dark energy models.

A second point which justifies the basic assumptions of the model is that the linear relationship between $\bar{\xi}$ 
and $\delta$ holds for virialized objects regardless the background cosmology. 
Moreover, there are a few works which advocate the validity of the stable clustering assumption and the machinery 
related to it: 
\cite{Padmanabhan1996a} showed that the basic physics behind the non-linear scaling relations obeyed by the two-point 
correlation function can be obtained from an appropriate use of the spherical collapse model; \cite{Padmanabhan1996b} 
considered $\Lambda$CDM simulations and showed that stable clustering is a better approximation for models with 
$\Omega_{\rm m}<1$ in which structure formation freezes out at some low redshift, even if it is not universally true. 
Finally, \cite{Munshi1997} found good agreement between analytical predictions and numerical simulations using the 
stable clustering to describe $N$-point correlation functions.

Finally, it is worth to mention the work by Kanekar \cite{Kanekar2000}, published soon before \cite{Engineer2000}. In 
this work, the author shows that the existence of the stable clustering is related to the Davis-Peebles scale-invariant 
solution \citep{Davis1977}. Using the BBGKY moments \citep{Davis1977,Ruamsuwan1992,Yano1997} and the assumption that 
$h=h(\bar{\xi})$, \cite{Kanekar2000} showed that stable clustering is viable, but the standard picture $h\rightarrow1$ 
as $\bar{\xi}\rightarrow\infty$ is not correct, as a stability analysis rather requires $0\leq h\leq 1/2$. While this 
value is different from the one used by \cite{Engineer2000} and in this work, our analysis is not affected since the 
value of $q$ is determined on the peak of $h$, rather than on its asymptotic value. 
Further to this, \cite{Engineer2000} showed that their fit did not change by more than 1 percent, if the curves were 
constrained at the peak or at $\delta\approx 10^4$. 
We expect therefore that our results would not be affected by the exact choice of the asymptotic value of $h$. Even if 
this would be the case, we would expect at most a few percent change in the parameter $q$ and less in 
$\Delta_{\rm vir}$. Note also that, as long as the asymptotic value of $h$ is constant, the Taylor expansion of the 
$S(\delta)$ term is correct.

For a more in-depth discussion, we refer the interested reader to 
\cite{Davis1977,Ruamsuwan1992,Nityananda1994,Yano1997,Padmanabhan1997,Engineer2000,Kanekar2000}.

\section{Numerical solutions: comparison}\label{sect:analysis}

In this section, we analyse the predictions of the two approaches for the different cosmological models, firstly 
focusing on the EdS and $\Lambda$CDM models; see Appendix~\ref{sect:implementation} for the details of the numerical 
implementation.

Both approaches I and II rely on the peculiar velocity $h_{\rm SC}$ of the spherical shell, defined as
\begin{equation}\label{eqn:hsc}
 h_{\rm SC} \equiv \frac{1}{3}\frac{d\ln{(1+\delta)}}{d\ln{a}} = 1-\frac{R^\prime}{R}\,,
\end{equation}
which determines the value of the parameter $q$. 
As in \cite{Engineer2000,Shaw2008}, we will assume that $h_{\rm SC} = h_{\rm SC}(\delta)$, which from a numerical 
point-of-view is justified by the analysis of \cite{Hamilton1991}. The EdS model will allow us to test our general 
implementation with results obtained in \cite{Engineer2000} and \cite{Shaw2008}, while the $\Lambda$CDM model will 
serve as reference model to compare more general dark-energy models to.

\subsection{The virial term \texorpdfstring{$\boldsymbol{S(\delta)}$}{Sdelta}}

\begin{figure}
 \centering
 \includegraphics[scale=0.8]{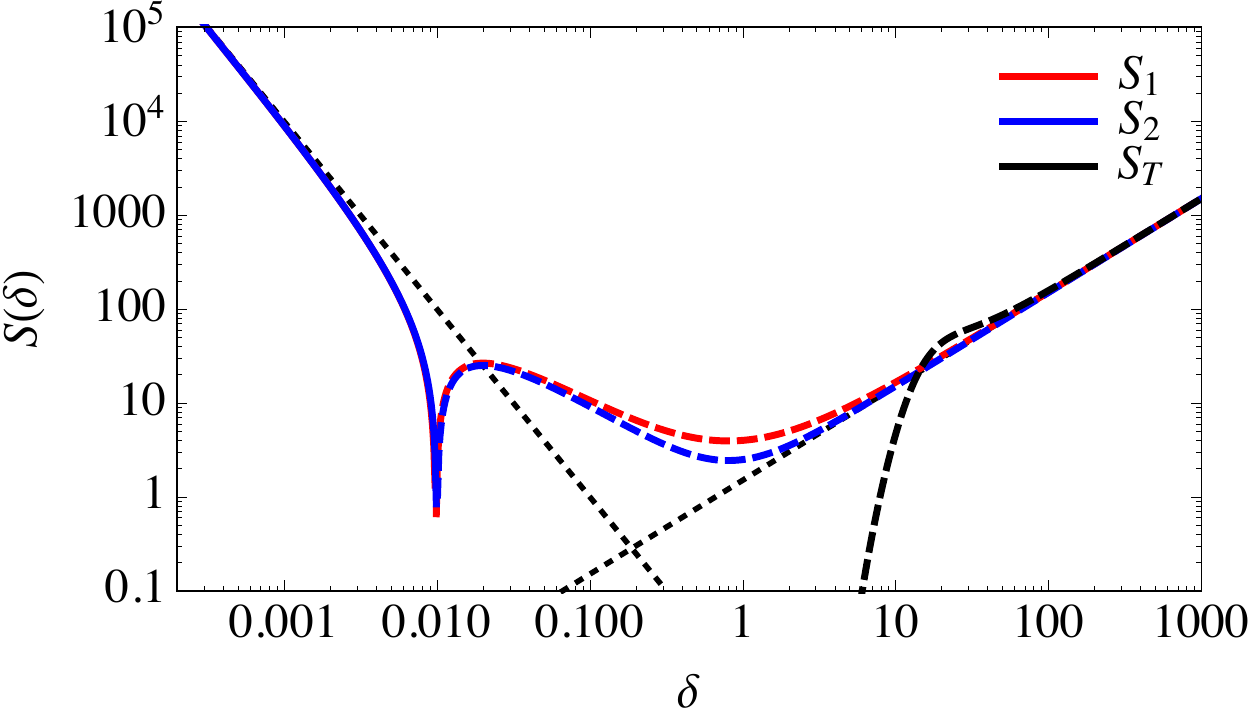}
 \caption{Evolution of the virialization term $S({\delta})$ as a function of $\delta$ for an EdS model (absolute value; 
 dashed lines correspond to negative values). The curves correspond to Equations~(\ref{eqn:S12}) with $q=0.01$ (red and 
 blue; \protect\cite{Engineer2000}) and Equation~(\ref{eqn:ST}) (black; \protect\cite{Shaw2008}). The straight dotted 
 lines represent the asymptotic regimes $q/\delta^2$ and $3\delta/2$: larger values of $q$ affect the shear-rotation 
 virialization term only in regions with small overdensity, i.e., in the linear regime.}
 \label{fig:Sdelta}
\end{figure}

As one can expect from the previous discussion in Section~\ref{sect:SCM}, the two models S$_1$ and S$_2$ and model 
S$_{\rm T}$, respectively Eqs.~(\ref{eqn:S12}) and~(\ref{eqn:ST}), must match for large values of the overdensity, 
$\delta\gg1$. This qualitative expectation is confirmed as shown in Figure~\ref{fig:Sdelta}, where we show the 
evolution of the virialization term $S(\delta)$ as function of $\delta$ for the EdS model. For models S$_1$ and S$_2$ 
we set $q=0.01$, a value which well approximates the $N$-body expectations for $h_{\rm SC}$. Similar results would be 
obtained for different values or for more generic models \citep{Engineer2000}, which would eventually only alter the 
slope for $\delta\ll1$ and the value of the overdensity $\delta_h$ for which the models S$_1$, S$_2$, and S$_{\rm T}$ 
do match.

As in \cite{Engineer2000}, the parameter $q$ for models S$_1$ and S$_2$ is determined by fitting the maximum of 
$h_{\rm SC}$ to appropriate $N$-body results. One should therefore use simulations with the correct cosmology, however 
the fitting prescription is available for only EdS simulations. One could instead determine $h_{\rm SC}$ and its 
amplitude from the expressions presented in \cite{Shaw2008}, but this would imply mixing the two procedures. In the 
following, when dealing with a generic dark-energy model, we shall fit for $q$ by matching the maximum of $h_{\rm SC}$ 
as provided by the EdS $N$-body simulation. 
As we will see in Section~\ref{sect:DeltaV}, even larger variations of $q$ within a factor of 2 would not lead to 
significant quantitative changes to the observed quantities.

How much the two recipes affect $\Delta_{\rm vir}$? As explained also in \cite{Engineer2000}, the virialization term 
is valid only for $\delta\gtrsim 15$. Models S$_1$ and S$_2$ essentially differ only for 
$0.1\lesssim\delta\lesssim 10$; see Figure~\ref{fig:Sdelta}. Even if the difference is small and might look marginal, 
it actually implies a value for $\Delta_{\rm vir}$ for the model S$_1$ about 6 percent larger than for model S$_2$. As 
for the model S$_{\rm T}$, $S(\delta)$ is appreciably different from zero only for $\delta\gtrsim10$, which implies a 
very small value for $\Delta_{\rm vir}$, about a factor 2.5 smaller than for model S$_1$. The reason is that the 
function $T[\eta(\tau)]\equiv T(\tau)\approx\tau$ and then $1/{T_\tau}^2-1\approx0$, except for high densities where 
it grows exponentially over a short range of time $\tau$.

It is now easy to understand why the term $S(\delta)$ acts as a virialization term: in the linear regime it is either 
vanishing or negative and counteracts the leading term originating from the Poisson equation. 
As long as $\delta\gtrsim 10$ the shear-rotation term and the gravitational force balance hampering the growth of 
perturbations, further impeded by the cosmological expansion.

\subsection{Evolution of the normalised radius \texorpdfstring{$\boldsymbol{y}$}{y}}\label{sect:radius}

We can restrict the discussion to the EdS and $\Lambda$CM models since the normalised radius $y$ evolves with time 
qualitatively in the same way in more general dark-energy models, with differences only in the amplitude and location 
of its maximum.

\begin{figure}
 \centering
 \includegraphics[scale=0.54]{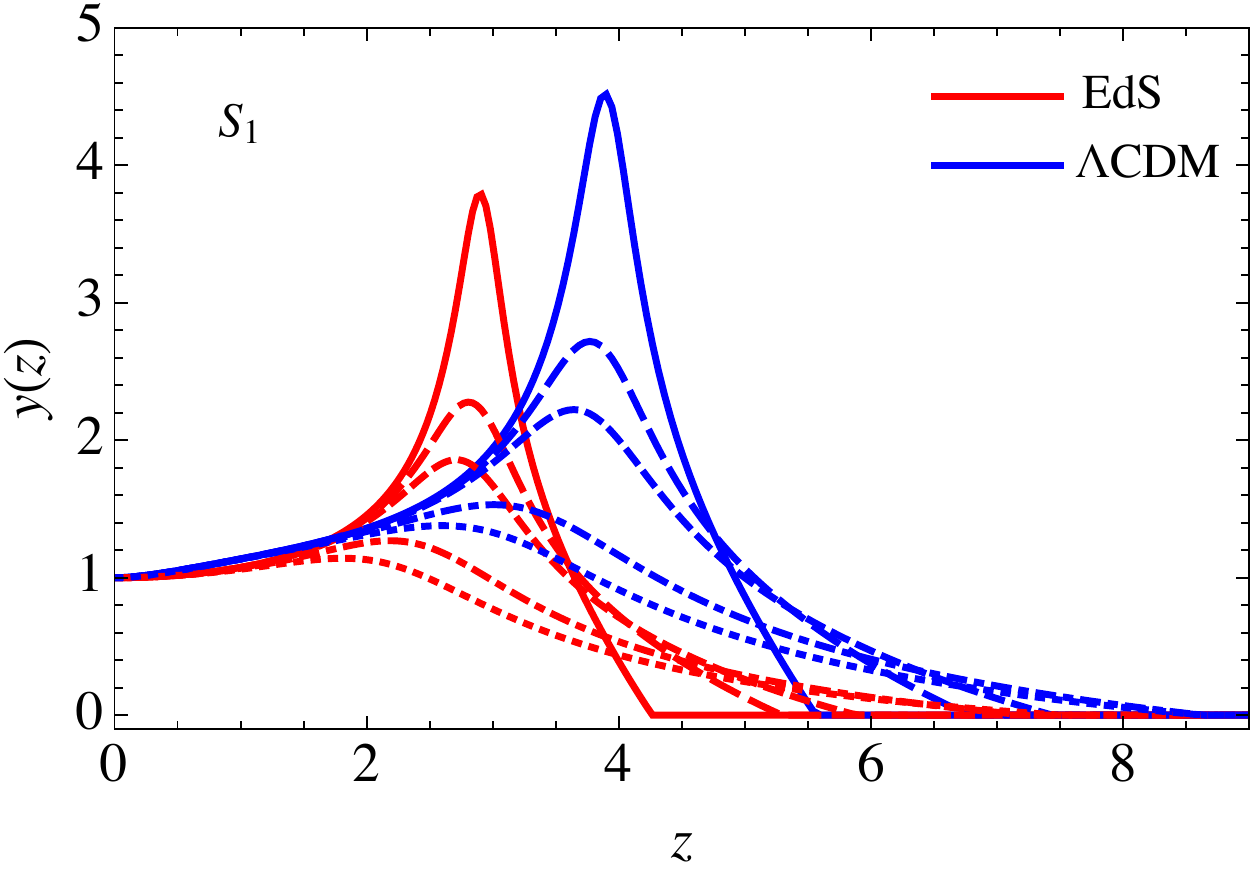}
 \includegraphics[scale=0.54]{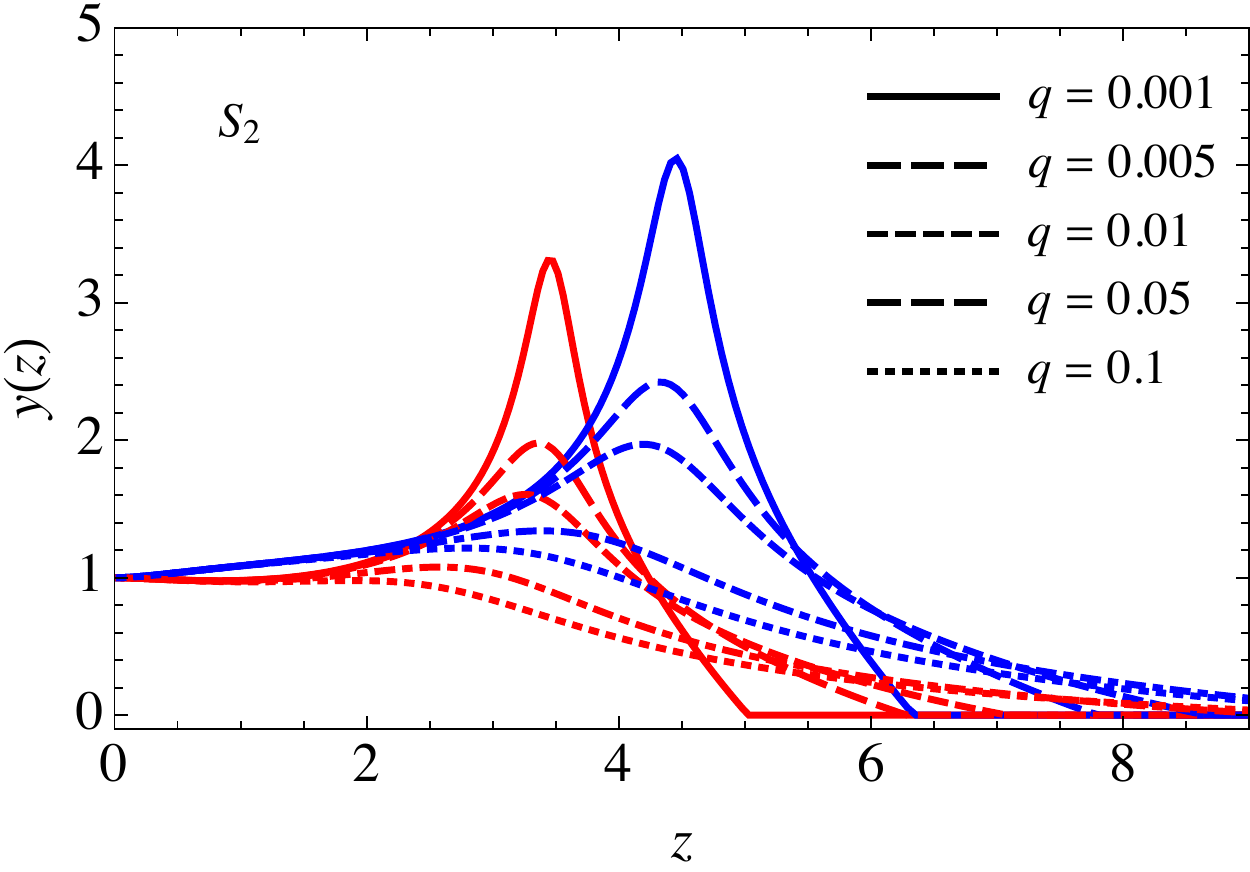}
 \caption{Evolution of the normalised radius $y$ as a function of the redshift $z$ for the EdS (red curves) and the 
 $\Lambda$CDM (blue curves) models for virialization models S$_1$ (left panel) and S$_2$ (right panel). Different line 
 styles correspond to different values of the parameter $q$, ranging from 0.001 to 0.1, as indicated.}
 \label{fig:yq_EdS_LCDM}
\end{figure}

Figure~\ref{fig:yq_EdS_LCDM} illustrates the evolution of $y(z)$ as a function of the redshift for models S$_1$ and 
S$_2$ (left and right panel, respectively). By increasing $q$ the maximum radius has a smaller value and is attained 
at later time. This happens because the term proportional to $y^7$ becomes more important, forcing the perturbation to 
decouple from the Hubble flow and reach the equilibrium at earlier time. At late times all the curves converge to one 
as expected, regardless of the value of $q$. Though counterintuitive, this result is a direct consequence of the 
adopted initial conditions which have been set at the time of collapse, here $z=0$. Setting the collapse for the 
standard SCM at earlier times simply shifts the turn-around at earlier times, without changing the overall picture. 
Moreover, comparing the left and right panels, the turn-around in model S$_2$ is anticipated with respect to model 
S$_1$ and the radius has a lower value. Indeed, in model S$_2$ there is an additional linear term $-\Omega_{\rm m}y/2$ 
which opposes the friction term proportional to $y^{\prime}$; see Eqs.~(\ref{eqn:yq}).

This behaviour is qualitatively unchanged for EdS and $\Lambda$CDM models, and will be similar for more general dark 
energy models. The turn-around occurs earlier for a $\Lambda$CDM model rather than for an EdS because of the same 
initial conditions, which fix the amplitude of the parametric solutions $R(\eta)$ and $t(\eta)$. Thus 
$H_{\Lambda\rm{CDM}}<H_{\rm EdS}$ and therefore the turn-around happens earlier. This is true in general for any dark 
energy model, as $T(\tau)\approx\tau$ for most of the cosmic evolution. At the same time, since 
$GM_{\rm m}\propto R^3/t^2$, the mass conservation implies $R_{\Lambda\rm{CDM}}>R_{\rm EdS}$ and therefore a higher 
value for the turn-around radius.

\begin{figure}
 \centering
 \includegraphics[height=4.7cm]{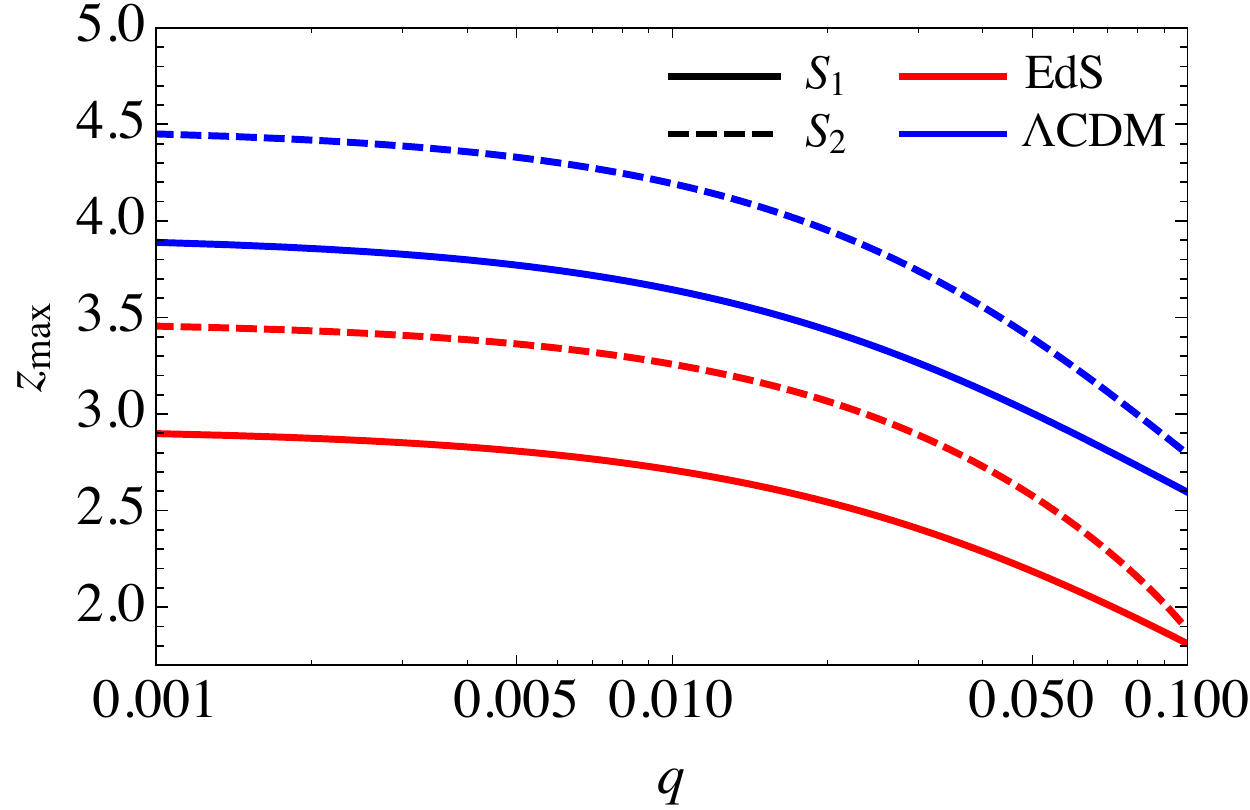}
 \includegraphics[height=4.7cm]{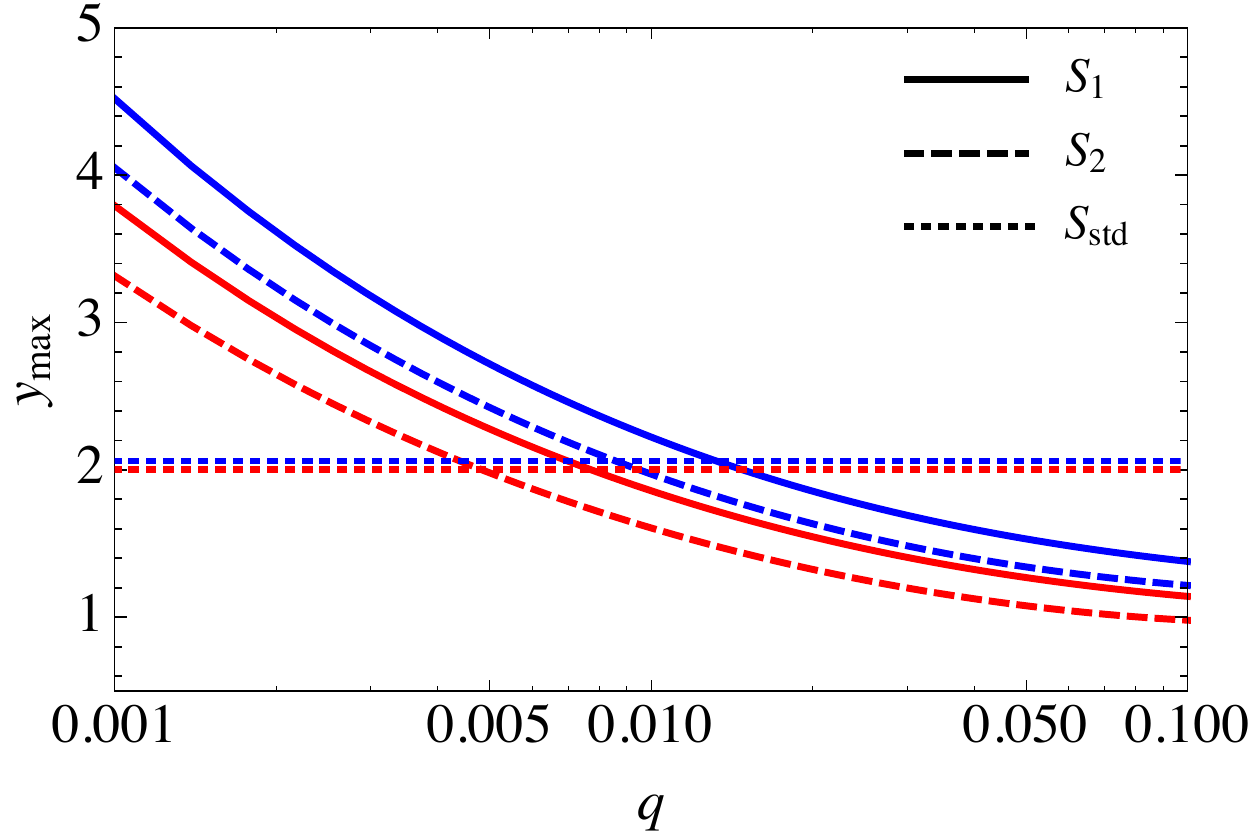}
 \caption{Epoch of maximum expansion $z_{\rm max}$ (left panel) and corresponding maximum normalised radius 
 $y_{\rm max}$ (right panel) as a function of the parameter $q$ for the EdS (red) and $\Lambda$CDM (blue) models. 
 Solid (dashed) lines refer to model S$_1$ (S$_2$). 
 The red (blue) dotted line shows the value of $y_{\rm max}$ for the standard virialization recipe (S$_{\rm std}$) 
 based on the virial theorem as presented in \protect\cite{Wang1998}.}
 \label{fig:a-y_max}
\end{figure}

One can finally study the turn-around, specifically the maximum value $y_{\rm max} =r_{\rm ta}/r_{\rm vir}$ and epoch 
at which it occurs, $z_{\rm max}$, as function of the free parameter $q$; see Figure~\ref{fig:a-y_max}. According to 
Figure~\ref{fig:yq_EdS_LCDM}, $z_{\rm max}$ decreases with $q$, with larger values for model S$_1$ than for S$_2$ 
regardless of the background cosmological model. When dark-energy dominates, turn-around takes place earlier. Also the 
value of $y_{\rm max}$ is higher for model S$_1$ than S$_2$, but it decreases when $q$ increases.

We immediately notice that the range of $y_{\rm max}$ in the classical recipe \citep{Wang1998} is very limited, 
allowing one to assume the EdS value also for the $\Lambda$CDM model. Indeed, when the standard virialization condition 
is considered, we find $y_{\rm max}=2$ for the EdS model and $y_{\rm max}\approx 2.06$ for the $\Lambda$CDM model 
(dotted lines in Figure~\ref{fig:a-y_max}). In the model proposed by \cite{Engineer2000}, the variability of 
$y_{\rm max}$ as a function of $q$ between the two models examined is larger. To reproduce these values, 
$y_{\rm max}\approx 2$ can be recovered for $q\approx 0.01$ within a factor of two. This is broadly in agreement with 
what found by \cite{Engineer2000} and we refer to this work and to \cite{Shaw2008} for a more detailed discussion on 
the values found in $N$-body simulations.

\subsection{Evolution of the peculiar velocity of the shell}\label{sect:hsc}

\begin{figure}
 \centering
 \includegraphics[scale=0.57]{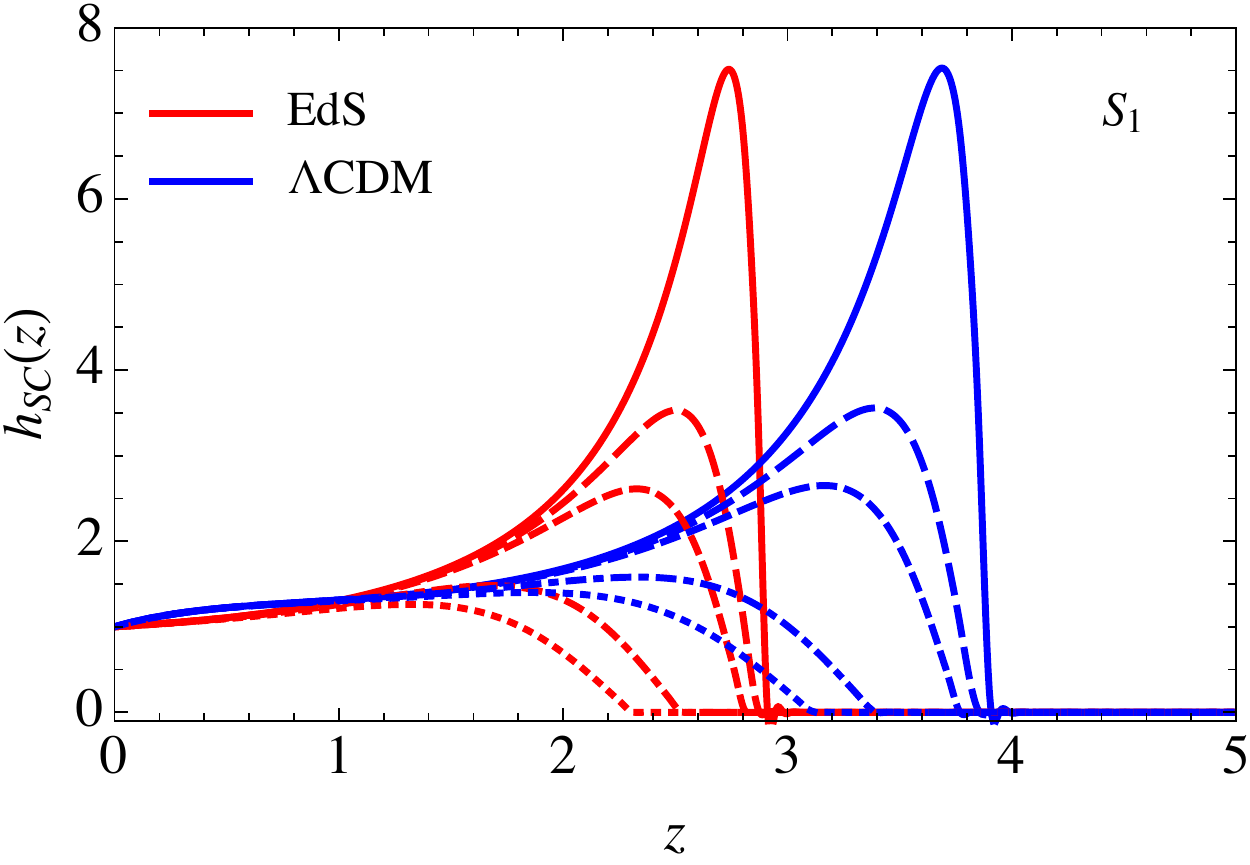}
 \includegraphics[scale=0.57]{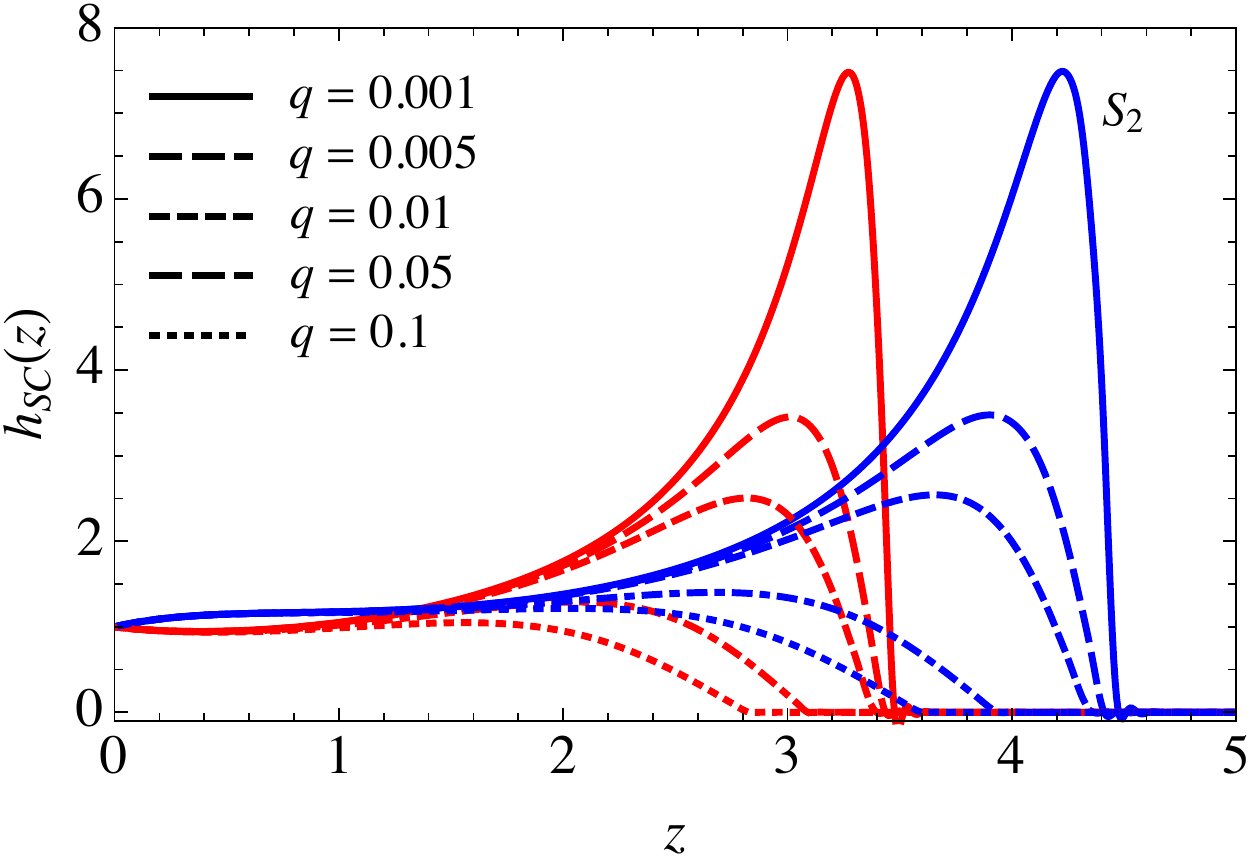}
 \caption{Evolution of the peculiar velocity of the shell, $h_{\rm SC}$, as a function of the redshift $z$ for the EdS 
 (red curves) and the $\Lambda$CDM (blue curves) models for virialization models S$_1$ (left panel) and S$_2$ (right 
 panel). Lines styles for different values of $q$ are as in Figure~\ref{fig:yq_EdS_LCDM}.}
 \label{fig:hsc_EdS_LCDM}
\end{figure}

The peculiar velocity of the shell $h_{\rm SC}$ deduced from the solution is studied as function of the redshift $z$ 
for several values of $q$ for the EdS and $\Lambda$CDM models; see Figure~\ref{fig:hsc_EdS_LCDM}. 
Consistently with the evolution of the normalised radius $y$, the maximum of $h_{\rm SC}$ is reached earlier in a 
$\Lambda$CDM model than in the EdS model and this is true for both models S$_1$ and S$_2$, the latter offering a 
slightly faster dynamics with anticipated onset of the accelerated phase.

Interestingly enough, the larger is $q$ the smaller is the maximum value of $h_{\rm SC}$. However, this value is 
largely insensitive to the background cosmological model: for small values of $q$ the peak has the same value for EdS 
and $\Lambda$CDM models and any difference between them arises only for relatively high values of $q$. As we will see 
in the next section, this suggests that the values of $q$ will be very similar for more general dark-energy models, 
justifying the use of an EdS simulation to fit the amplitude of the peculiar velocity also in different cosmologies.

A further justification comes from the virialization model proposed by \cite{Shaw2008}. 
Using the notation introduced in Section~\ref{sect:SCM:approach2}, one can show that
\begin{equation}
 h_{\rm SC} = 1 - \frac{3}{2}\sqrt{\Omega_{\rm m}(a)\tilde{f}(a)}\frac{\sin{\eta}}{(1-\cos{\eta})^2}
                  \frac{T(\eta)}{T_\tau(\eta)}\,.
\end{equation}
With $\tilde{f}(a)\approx[\Omega_{\rm m}(a)]^{-0.4}$, one obtains
\begin{equation}
 h_{\rm SC} \approx 1 - \frac{3}{2}[\Omega_{\rm m}(a)]^{0.3}\frac{\sin{\eta}}{(1-\cos{\eta})^2}
                        \frac{T(\eta)}{T_\tau(\eta)}\,.
\end{equation}
At $z=0$ and with $\Omega_{\rm m,0}\approx0.3$, the value of $q$ changes by less than a factor of 2, implying a change 
in $\Delta_{\rm vir}$ by less than a percent.

\subsection{Evolution of the virial overdensity in smooth dark-energy models}\label{sect:DeltaV}

\begin{figure}
 \begin{center}
  \includegraphics[scale=0.57]{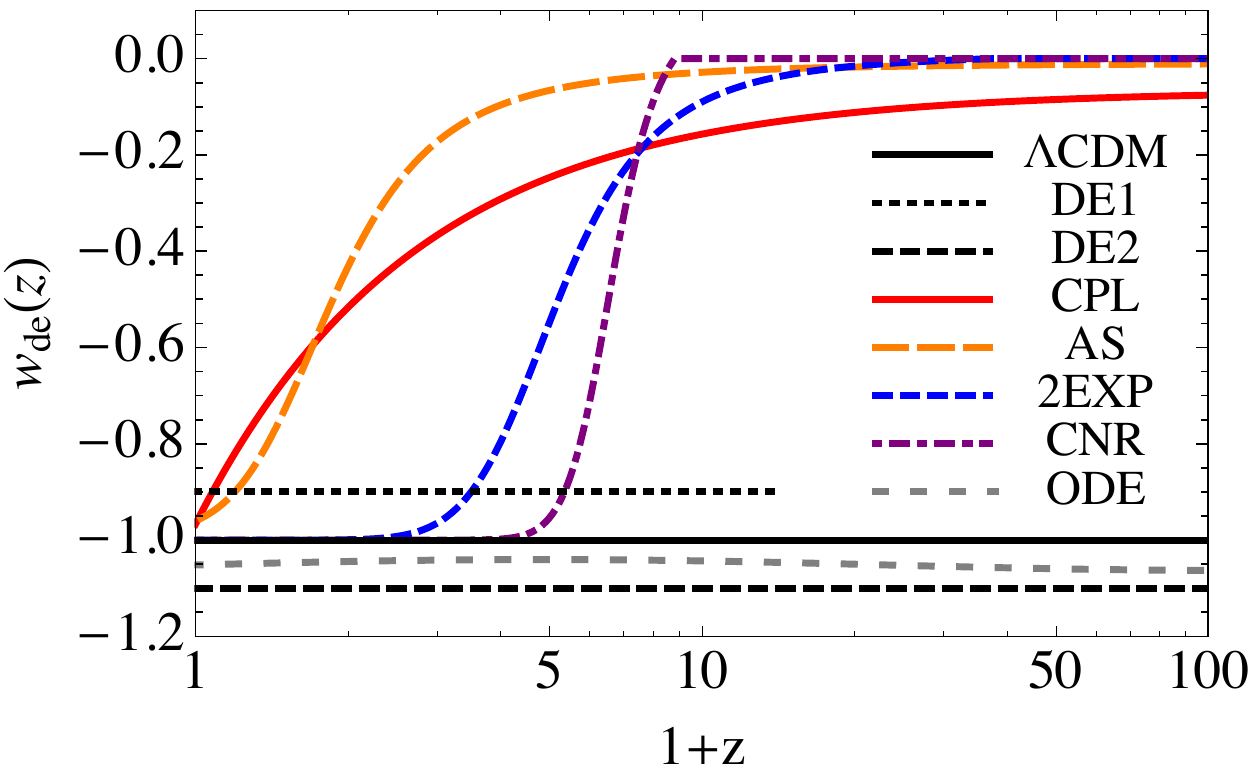}
  \caption{Time evolution of the equation-of-state $w_{\rm de}(z)$ as a function of the redshift $z$ for the 
  dynamical dark-energy models considered in this work. The three horizontal black lines represent models with constant 
  equation-of-state: DE1 with $w=-0.9$ (dotted black line), $\Lambda$CDM with $w=-1$ (solid black line), and DE2 with 
  $w=-1.1$ (short-dashed black line). The solid red (dashed orange) line shows the CPL (AS) model, the dashed blue 
  (violet dot-dashed) line shows the 2EXP (CNR) model, and the short dashed grey line represents the ODE model.}
  \label{fig:wa}
 \end{center}
\end{figure}

This section presents the main achievement of this work, namely the time evolution of the virial overdensity 
$\Delta_{\rm vir}$ in presence of dark-energy. Beside the EdS model, useful to validate our numerical implementation, 
and the $\Lambda$CDM model, used as reference model, we focus on seven specific dark-energy models that are compatible 
with $\Lambda$CDM today and cover a wide range of behaviour and represent a good test for the proposed virialization 
recipes. Their effects are limited to the background expansion and therefore fully described by the equation-of-state 
(see Figure~\ref{fig:wa}):
\begin{itemize}
 \item two dark-energy models with constant equation-of-state,
       \begin{equation}
        \hspace{0.7cm} w_{\rm de} = -0.9\;\mbox{(DE1),}\quad w_{\rm de} = -1.1\;\mbox{(DE2)}\,;
       \end{equation}
 \item the CPL model \citep{Chevallier2001,Linder2003}, with equation-of-state
       \begin{equation}
        \hspace{0.7cm} w_{\rm de}(a) = w_0 + w_a(1-a)\,,
       \end{equation}
       with $w_0=-0.967$ and $w_a=0.9$ \citep{Avsajanishvili2014};
 \item the oscillating model \citep[ODE;][]{Pace2012,Pan2017,Panotopoulos2018}, with equation-of-state
       \begin{equation}
        \hspace{0.7cm} w_{\rm de}(a) = w_0 + b\sin{[\ln{(1/a)}]}\,,
       \end{equation}
       with $w_0=-1.0517$ and $b=0.0113$;
 \item the CNR \citep{Copeland2000}, 2EXP \citep{Barreiro2000}, and AS \cite{Albrecht2000} models, which collectively 
       can be described by \citep{Corasaniti2003}
       \begin{equation}
        \hspace{0.7cm} w_{\rm de}(a) = w_0 + (w_{\rm m}-w_0)
        \frac{1+\mathrm{e}^{\frac{a_{\rm m}}{\Delta_{\rm m}}}}{1+\mathrm{e}^{-\frac{a-a_{\rm m}}{\Delta_{\rm m}}}}
        \frac{1-\mathrm{e}^{-\frac{a-1}{\Delta_{\rm m}}}}{1-\mathrm{e}^{\frac{1}{\Delta_{\rm m}}}}\,,
       \end{equation}
       with the same parameters used in \cite{Pace2010}; see Table~\ref{tab:param}.
\end{itemize}

\begin{table}
 \caption{Parameter values for the quintessence models.}
 \begin{center}
  \begin{tabular}{ccccc}
  \hline
  Model & $w_0$ & $w_{\rm m}$ & $a_{\rm m}$ & $\Delta_{\rm m}$ \\
  \hline
  2EXP  & -1.0  &  0.01 & 0.19  & 0.043      \\
  AS    & -0.96 & -0.01 & 0.53  & 0.13       \\
  CNR   & -1.0  &  0.1  & 0.15  & 0.016      \\
  \hline
  \end{tabular}
  \label{tab:param}
 \end{center}
\end{table}

The AS model departs very quickly from the value at $z\approx 0$ and grows asymptotically to $w=0$, while the CPL 
model, despite having the same qualitative behaviour, has a much gentler departure from its value today and reaches 
$w=-0.1$ at early times. The 2EXP and the CNR models have a rather flat equation-of-state at late times, closely 
resembling a cosmological constant, rapidly growing to zero in a redshift interval that might be as short as 
$\Delta z\sim3$. The ODE model has a very long period of oscillation, changing its value very slowly and looking 
approximately constant and in the phantom regime over the redshift range $0\leqslant z \lesssim 100$.

Since the qualitative evolution of the radius and peculiar velocity does not change with respect to the $\Lambda$CDM 
model, we limit the discussion to the redshift evolution of $\Delta_{\rm vir}$ and compare the standard results 
obtained by using the prescription by \cite{Wang1998,Maor2005} at the virialization time $z_{\rm vir}$ with those of 
\cite{Engineer2000} and \cite{Shaw2008}. For all three approaches, $\Delta_{\rm vir}=1+\delta_{\rm NL}$ where 
$\delta_{\rm NL}$ represents the non-linear evolution of the matter overdensity. Note that while for 
\cite{Wang1998,Maor2005} the integration is only up to $z_{\rm vir}$, for the latter we do carry the integration till 
the collapse time of the standard SCM.

To study the evolution of $\Delta_{\rm vir}$ in a given model, for every cosmology we first look for the value of the 
free parameter $q$ such that the maximum value of $h_{\rm SC}$ matches the value found from $N$-body simulations for 
the EdS model. Even though this procedure is in principle not consistent, it has a quantitative negligible impact on 
the final results. The values of $q$ reported in Table~\ref{tab:qpar} for both expressions S$_1$ and S$_2$ are clearly 
largely insensitive to the background cosmological model, with a maximum variation of about 8 percent, and all 
consistent with the value of $q\approx0.02$ given in \cite{Engineer2000}. As expected, values for the expression S$_2$ 
are smaller than those for S$_1$. We also note that this independence on the background cosmology reflects on the lack 
of any clear trend distinguishing quintessence from phantom models.

\begin{table}
 \caption{Values of the free parameter $q$ for the virialization model of \protect\cite{Engineer2000} for the EdS, 
 $\Lambda$CDM and the dark-energy cosmologies studied in this work using the virialization terms S$_1$ and S$_2$.}
 \begin{center}
  \begin{tabular}{ccc}
  \hline
  Model        & $q_{\rm S_1}$ & $q_{\rm S_2}$\\
  \hline
  EdS          &    0.0177    &    0.0151   \\
  $\Lambda$CDM &    0.0190    &    0.0160   \\
  DE1          &    0.0191    &    0.0161   \\
  DE2          &    0.0189    &    0.0159   \\
  CPL          &    0.0184    &    0.0160   \\
  2EXP         &    0.0189    &    0.0159   \\
  AS           &    0.0178    &    0.0156   \\
  CNR          &    0.0190    &    0.0160   \\
  ODE          &    0.0189    &    0.0160   \\
  \hline
  \end{tabular}
  \label{tab:qpar}
 \end{center}
\end{table}

\begin{figure*}
 \centering
 \includegraphics[scale=0.27]{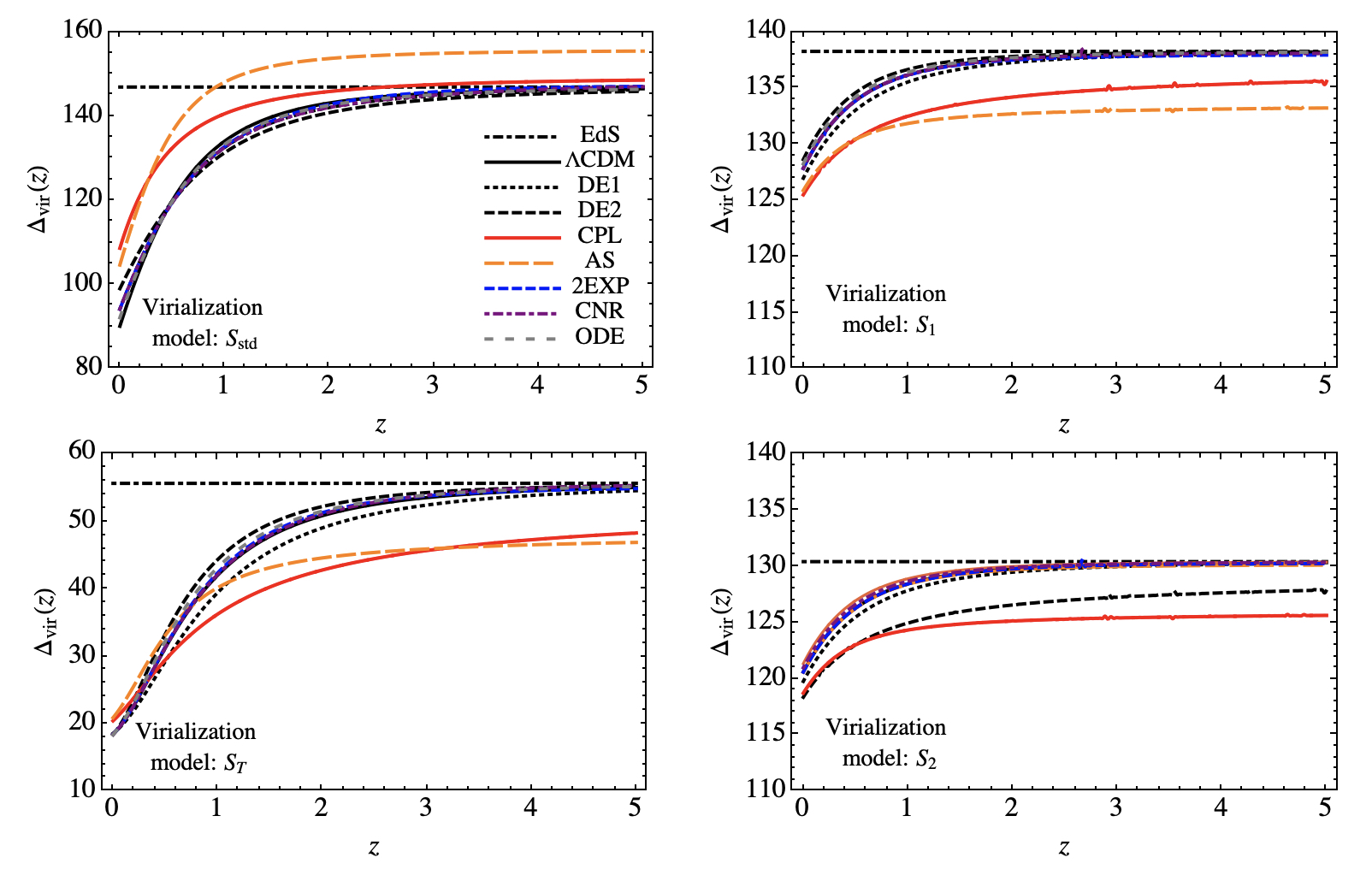}
 \caption{Evolution of the virial overdensity $\Delta_{\rm vir}$ as a function of redshift $z$ for different
 virialization recipes and dark-energy models (see legend). Clockwise from \textit{top left:} standard model 
 S$_{\rm std}$ by \protect\cite{Wang1998} and \protect\cite{Maor2005}, evaluated at the virialization redshift. 
 \textit{Top and bottom right:} models S$_1$ and S$_2$ derived from \protect\cite{Engineer2000}, respectively. 
 \textit{Bottom left:} virialization model by \protect\cite{Shaw2008}.}
 \label{fig:DeltaV}
\end{figure*}

The values of $q$ are used to calculate the evolution of $\Delta_{\rm vir}$ as a function of the collapse redshift 
$z_{\rm c}$. Note that for the standard prescription, $\Delta_{\rm vir}$ is evaluated at the virialization redshift 
$z_{\rm vir}$ corresponding to the collapse $z_{\rm c}$. The results are presented in Figure~\ref{fig:DeltaV}, in which 
each panel accounts for a different recipe for the virialization; clockwise from the top-right panel the \textit{a 
posteriori} virialization recipe by \cite{Wang1998,Maor2005}, the shear-rotation--induced parametric models S$_1$ and 
S$_2$ from \cite{Engineer2000}, and the S$_{\rm T}$ model from \cite{Shaw2008}. The qualitative behaviour of the three 
virialization recipes S$_1$, S$_2$, and S$_{\rm T}$ is the same for all the dark-energy models, however quantitative 
differences exist.

If the virialization is forced \textit{a posteriori} (top-left panel), all the dark-energy models but CPL and AS 
resemble very closely to the $\Lambda$CDM. This is explained in terms of the equation-of-state $w_{\rm de}$, which for 
all but the CPL and AS models closely follows the $\Lambda$CDM value $-1$ until redshift $z\simeq1.5-4$, while CPL and 
AS equations-of-state rapidly change in the late-time universe. It is worth to notice that among all the models with 
evolution very similar to $\Lambda$CDM, the models DE1 and DE2 maximally deviate. For $w_{\rm de}=-0.9$ the virial 
overdensity is smaller than for the $\Lambda$CDM model, which in turn is smaller than for the cosmology with 
$w_{\rm de}=-1.1$. 
This happens because decreasing the value of $w_{\rm de}$, the contribution of the dark-energy becomes more important 
and initial overdensities need to be larger to allow structures to collapse. These results are in agreement with 
\cite{Pace2010,Pace2012}, where the virial overdensity was evaluated at the collapse time; instead, here we consider 
$\Delta_{\rm vir}$ at the virialization redshift to allow for a direct comparison with \cite{Engineer2000} and 
\cite{Shaw2008}.

When virialization is induced by shear and rotation and models S$_1$ and S$_2$ are used (top and bottom right panels), 
the qualitative results are similar. For the EdS model, $\Delta_{\rm vir}$ is essentially constant with redshift, with 
differences smaller than 0.01 percent that are very likely of numerical and not physical origin. Again, all but CPL 
and the AS models approach the EdS value at high redshift. Quantitatively, according to \cite{Engineer2000} the 
virialization overdensity is lower by about 10 percent. Moreover, the values $\Delta_{\rm vir}$ for the recipe S$_1$ 
(top-right panel) are larger than for S$_2$ (bottom-right panel) because $S_2(\delta)<S_1(\delta)$ when $\delta\sim1$, 
as shown in Figure~\ref{fig:Sdelta}. This also translates into a smaller variation between low and high redshifts: for 
the standard approach, variations in time are of the order of 40 percent, while for \cite{Engineer2000} it was not 
larger than 10 percent.

Finally, with virialization inferred by $N$-body simulations, model S$_{\rm T}$ (bottom left panel), the values for 
$\Delta_{\rm vir}$ are overall much lower and the variability between low and high-redshift is much higher, about 60 
percent. At low redshift $\Delta_{\rm vir}\approx 20$ for the S$_{\rm T}$ model and $\approx 90-120$ for the three 
other virialization recipes; at high redshift, $\Delta_{\rm vir}\approx 55$ for S$_{\rm T}$ and $\approx 130-150$ for 
the other recipes. This is explained noting that $S(\delta)$ becomes important only for relatively high over-densities, 
forcing the non-linearities to kick in at later stages of the evolution of the perturbations.

We conclude this section by analysing more in detail the issue of the validity of the equations of motion used in our 
work. The models we considered here are purely phenomenological, i.e., not derived from a Lagrangian. 
Without looking for exotic models, it is nonetheless possible to derive the equations of state above using the 
formalism outlined in \cite{Battye2016}. For quintessence and $k$-essence Lagrangians, the models analysed in this 
work will still satisfy the Birkhoff's theorem and shells evolve independently. This is the case indeed, as these 
models do not need any screening mechanism to be viable in the Solar System and their Lagrangian does not change the 
laws of gravity. On the contrary, this would not be the case for other kind of models, such as KGB 
\citep{Deffayet2010,Pujolas2011,Kobayashi2010,Kimura2011} and $f(R)$ 
\citep{Silvestri2009,Sotiriou2010,DeFelice2010} cosmologies. In fact, in quintessence and $k$-essence models the 
effective gravitational constant $G_{\rm eff}$ is unchanged \cite{Pace2019}. The validity of the approach followed 
here is further justified by previous work \citep{Pace2010}, where it has been shown that even for early dark energy 
models the analytical treatment gives an excellent agreement with $N$-body simulations. 
We further stress that the key-point is the absence of a screening mechanism. In fact, Brans-Dicke models 
\citep{Brans1961} present both anisotropic stress and a time- but not scale-dependent gravitational constant and do 
not require any screening mechanism to be viable in the Solar System. This is achieved by requiring a (constant) large 
Brans-Dicke parameter, $\omega_{\rm BD}\gg 1$. In a more general settings, Ref.~\cite{Pace2014} considered a model 
largely studied in the literature, where $\omega_{\rm BD}=\omega_{\rm BD}(\phi)$, i.e., a function of the quintessence 
scalar field. As there is no screening, the equation of motion for the non-linear evolution of density perturbations, 
Eq.~(\ref{eqn:nldelta}), is simply replaced by the appropriate $G_{\rm eff}$ like in Eq.~(39) of Ref.~\cite{Pace2014}. 
The authors showed there that the spherical collapse formalism suitably modified by taking into account this aspect 
well reproduces the outcome of hydrodynamical simulations.

An alternative interesting point-of-view, also investigated in \cite{Lue2004}, is worth to be mentioned. 
It can be shown that the formalism is valid and the equations correct, provided that a Schwarzschild-like solution to 
Einstein field equations exists, and that the validity or the violation of Birkhoff's theorem can be used as a 
discriminant between dark energy and modified gravity to explain the late time accelerations. Though, note that the 
authors argue that their conclusions might be more generic and perhaps be independent from the validity of Birkhoff's 
theorem.

\subsection{Remark on the mass function}

Besides affecting the dynamics of single haloes, these new recipes shall affect also their number counts, i.e., the 
mass function. To identify haloes, both in $N$-body simulations and observations one usually employs Friend-of-Friend 
\citep[FoF;][]{Huchra1982,Einasto1984,Davis1985} or spherical overdensity \citep[SO;][]{Lacey1994,Eke1996a,White2001} 
algorithms, which in general yield different results depending on the range of masses and redshifts and eventually on 
cosmology \citep{White2001}. By construction, the FoF halo mass is larger than the SO halo mass, which is defined 
according to some mean overdensity $\Delta$ almost proportional to $\Delta_{\rm vir}$ 
\citep[see appendix C of][]{Hu2003}. While this procedure (choice of FoF vs. SO) has negligible effects for 
high-mass haloes, it usually wipes out low-mass haloes. If one of the parametric virialization models S$_1$ or S$_2$ is 
correct, then a lower value of $\Delta$ should be consistently considered; the haloes will therefore be spatially more 
extended, with possible consequence on the low-mass end of the mass function.

Since the values found for $\Delta_{\rm vir}$ are quite different for the recipes investigated here and the 
definition of haloes heavily rely on this quantity, it is important to study how the mass function will be affected. 
While this will not inform much about the nature of dark energy (though some model such as AS and CPL could be 
excluded as problematic at high redshift) as this would require studying the virialization also in modified gravity 
models, it could instead provide fitting formulae for an improved halo mass function as a function of the virial 
overdensity. To achieve this goal, one should construct halo catalogues according to a given value of 
$\Delta_{\rm vir}$ and fit the numerical mass function, whose free parameters will depend on the virial overdensity 
and ultimately on the prescription used. This goes beyond the purpose of this work and we will not perform this study 
in this work.

\begin{figure*}
 \centering
 \includegraphics[scale=0.21]{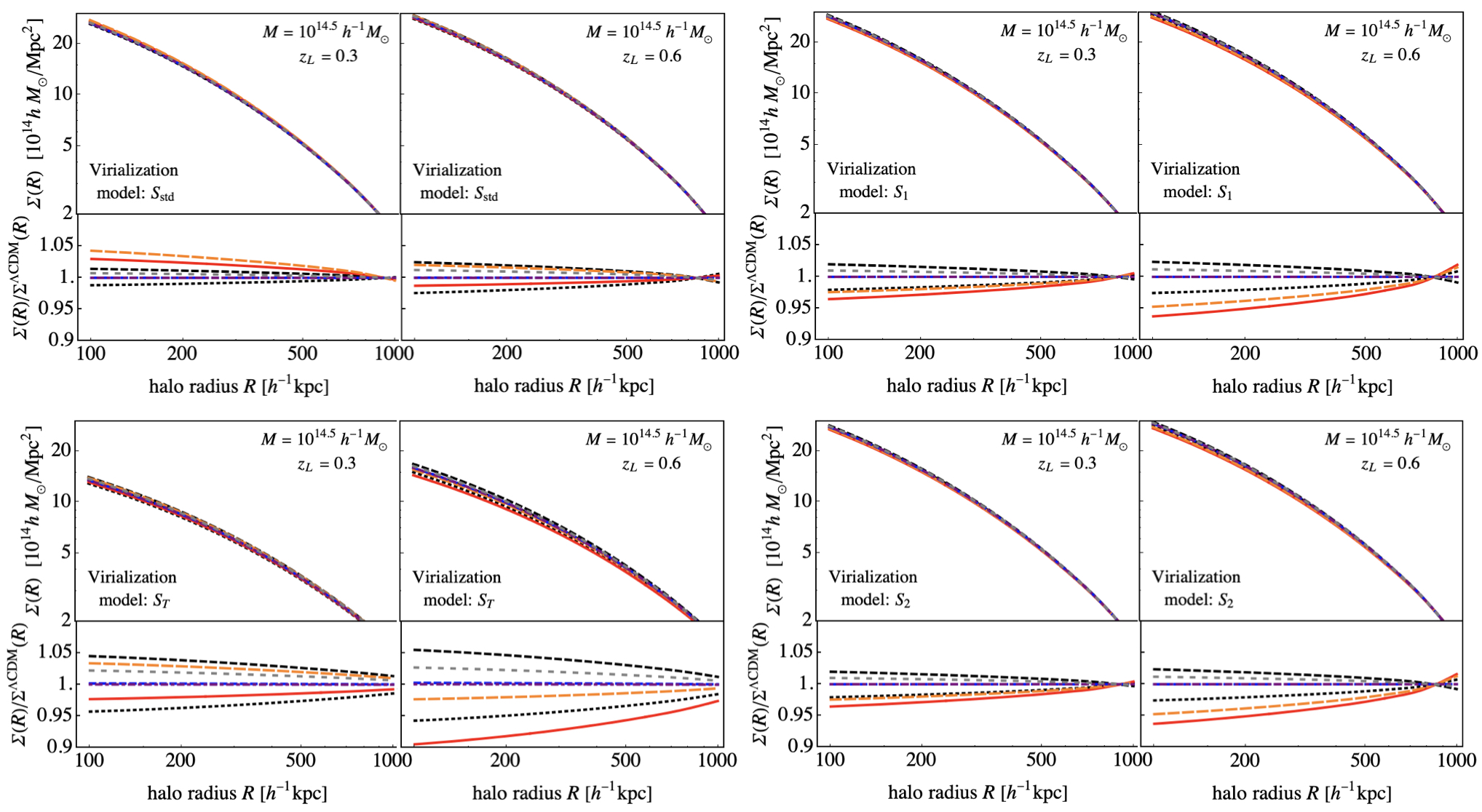}
 \caption{Radial surface-mass-density of dark-matter halo with virial mass $M=10^{14.5}h^{-1}M_\odot$ and truncated 
 NFW profile at redshift $z_{\rm L}=0.3$ and $z_{\rm L}=0.6$ (left and right panel of each box), for different 
 virialization recipes and dark-energy models as indicated (same legend as in Figure~\ref{fig:DeltaV}). The 5--10\% 
 deviations from the $\Lambda$CDM model change with the redshift of the halo depending on the virialization recipe, as 
 expected, and are almost insensitive on the halo mass and the concentration parameter.}
 \label{fig:SigmaWL}
\end{figure*}

\section{Weak-lensing observables}\label{sect:lensing}
Local observables that are essentially not related to the large-scale structure as uncoupled from its dynamics, such as 
the surface-mass-density profile of dark matter haloes measured from gravitational (weak) lensing, can provide a more 
secure insight into the virialization recipes in the different dark-energy scenarios. For illustrative purpose, we 
considered spherically averaged dark-matter haloes with a NFW profile \citep{Navarro1997} with mass- and 
redshift-dependent concentration parameter according to \cite{Duffy2008}. The surface mass density $\Sigma(R)$ 
projected along the line-of-sight is calculated as a function of the radius of the halo, sharply truncated at the 
virial radius as in \cite{Takada2003b}. This crude approximation can be improved by smooth truncation accounting for 
tidal effects \citep[e.g.][]{Baltz2009}, however it goes beyond the purpose of this paper. The results for the 
different virialization and dark-energy models are illustrated in Figure~\ref{fig:SigmaWL} for haloes of virial mass 
$M=10^{14.5}h^{-1}M_\odot$ at redshift $z_{\rm L}=0.3$ and $z_{\rm L}=0.6$, with concentration parameter at 
virialization radius respectively $c_\mathrm{vir}=4.3$ and 3.7. The relative differences are insensitive to the mass of 
haloes until $10^{16}h^{-1}M_\odot$ and to values of the concentration parameter in the range $c_\mathrm{vir}=3-9$. 
Interestingly enough, consistently with the redshift dependence of $\Delta_{\rm vir}(z)$, the deviation from the 
$\Lambda$CDM depends on the redshift of the halo in a non-trivial way. These deviations are degenerate with the 
virialization model, the models S$_1$ and S$_2$ being not distinguishable at this level while the model S$_\mathrm{T}$ 
yielding the strongest deviations.

\begin{figure*}
 \centering
 \includegraphics[scale=0.21]{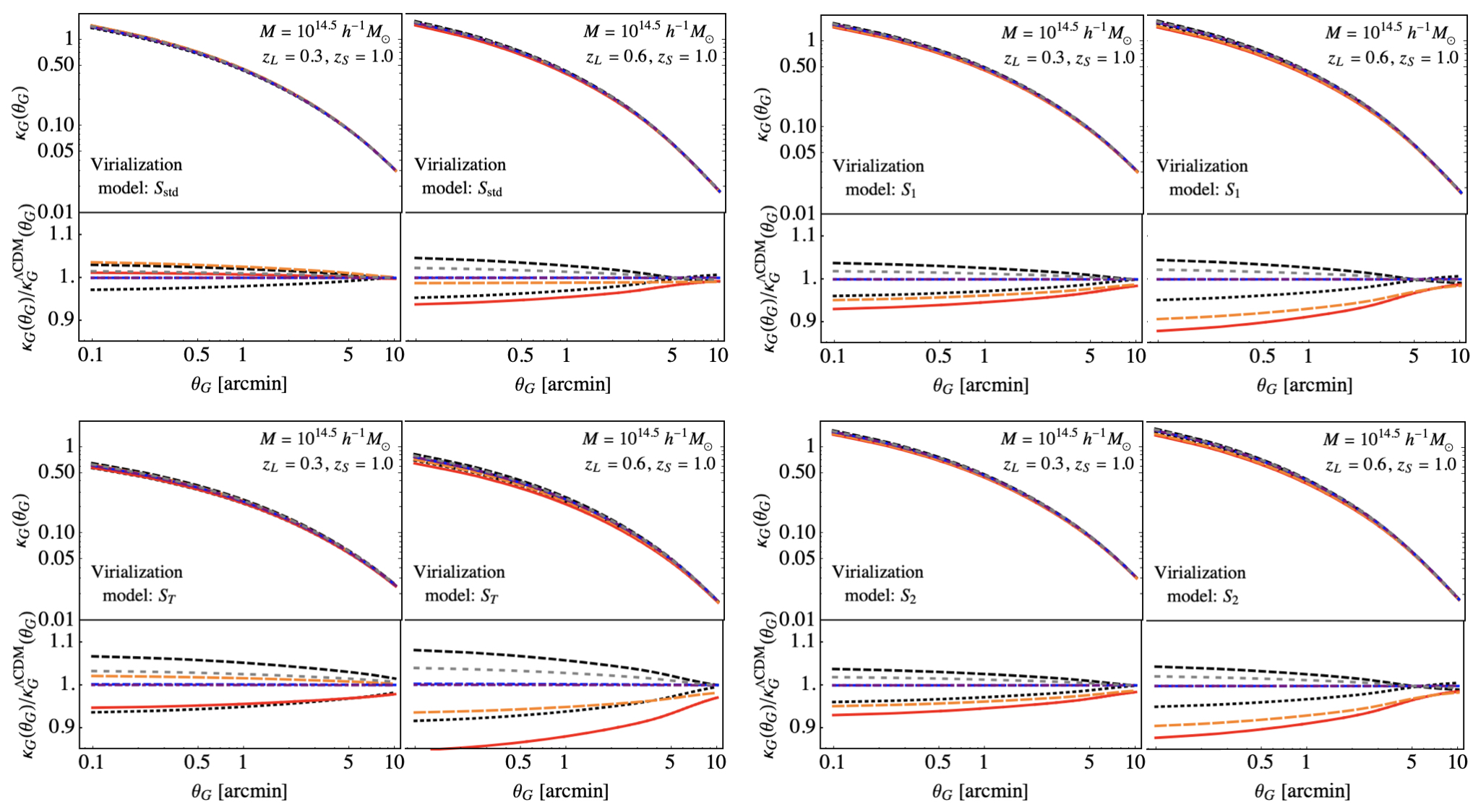}
 \caption{Gaussian-filtered convergence as function of the smoothing angular scale $\theta_{\rm G}$ for dark-matter 
 halo with virial mass $M=10^{14.5}h^{-1}M_\odot$ and truncated NFW profile at redshift $z_{\rm L}=0.3$ and 
 $z_{\rm L}=0.6$ and sources at redshift $z_{\rm S}=1$, for different virialization recipes and dark-energy models as 
 in Figure~\ref{fig:SigmaWL}.}
 \label{fig:kappaGsinglesourceDE}
\end{figure*}

\begin{figure}
 \centering
 \includegraphics[width=8cm]{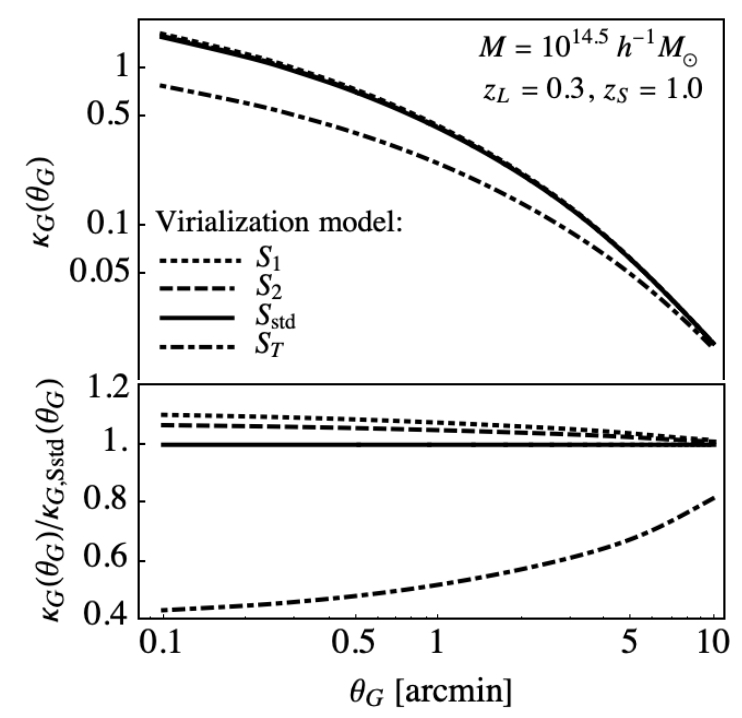}
 \caption{Virialization models in $\Lambda$CDM cosmology. Gaussian-filtered convergence as function of the angular 
 scale for a dark-matter halo with virial mass $M=10^{14.5}h^{-1}M_\odot$ and truncated NFW profile 
 ($c_\mathrm{vir}=4.3$) at redshift $z_{\rm L}=0.3$ and lensed sources at redshift $z_{\rm S}=1$. 
 The alternative dark-energy models share the same qualitative trend as function of $\theta_G$.}
 \label{fig:kappaGsinglesourceLCDM}
\end{figure}

Similar degeneracies between the dark-energy and virialization models persist when considering the (weak) lensing 
convergence field of haloes as function of the angular distance from the centre of the halo,
\begin{equation}
 \kappa(\theta) = \frac{\Sigma}{\Sigma_\mathrm{cr}} 
               \equiv \frac{4\pi G}{c^2}\frac{D_{\rm L}D_{\rm LS}}{D_{\rm S}}\Sigma(R=D_{\rm L}\theta)\,,
\end{equation}
the critical surface mass density $\Sigma_\mathrm{cr}$ further introducing a difference among the dark-energy models 
characterised by different angular distances $D_{\rm L}$, $D_{\rm S}$, and $D_{\rm LS}$. For a more realistic 
description, following \cite{Hamana2004} we consider the azimuthally averaged convergence 
$\kappa_G=\int\mathrm{d}^2\boldsymbol{\theta}W(\boldsymbol{\theta};\theta_G)\kappa(\theta)$ with the centre of the 
smoothing kernel $W$ set to the halo centre; see Figure~\ref{fig:kappaGsinglesourceDE}. For all the dark-energy models 
the relative difference from the $\Lambda$CDM is about 5-10 percent, decreasing for larger smoothing angles, and 
typically larger for the non-standard virialization models S$_1$, S$_2$ and S$_{\rm T}$, the latter yielding the 
largest deviation, as expected. For all but the CPL and AS dark-energy models these differences only slightly increase 
for haloes at larger redshift up to $z\lesssim 1$. Instead, because of their overall lowest convergence, the CPL and 
AS models can deviate by more than 30 percent at small angular scales with respect to the $\Lambda$CDM, moreover with 
a peculiar dependence on the redshift of the halo. It is worth to notice that the deviations among the dark-energy 
models are of the same order of the deviation between the virialization models. Focusing on $\Lambda$CDM and using the 
S$_{\rm std}$ as reference (Figure~\ref{fig:kappaGsinglesourceLCDM}), these are of order of 5-10 percent close to the 
halo centre, with a mild distinction between S$_1$ and S$_2$, and larger than 30 percent for the S$_{\rm T}$ model.

\subsection{Abundance of convergence peaks}

The results of the last section are used to estimate the number counts of convergence peaks measured by 
\textit{LSST}-like and 
\textit{Euclid}-like weak-lensing surveys. For this purpose, we calculated the halo mass function 
$n(M)=\frac{\rho_\mathrm{m}}{M^2}\frac{{\rm d}\log\sigma^{-1}}{{\rm d}\log M}f(\sigma)$ using the linear rms-mass 
variance $\sigma\equiv\sigma(M,z)$ and the multiplicity function $f(\sigma)$ of \cite{Watson2013} based on the FoF 
algorithm (see their Table 2, first column), thus not relying on any identification of haloes based on spherical 
overdensity. Although this multiplicity function has been fitted on a $\Lambda$CDM $N$-body simulation, we expect that 
the cosmological (dark-energy) model is essentially encoded in the redshift dependence of $\sigma$, which scales with 
the linear growth factor of the appropriate dark-energy model. The two major caveats of our computation are the mass 
dependence of $\sigma$, for which we assume a $\Lambda$CDM power spectrum, and the limitation of the FoF algorithm, 
which is influenced by the mass resolution and the linking length especially at high mass while only the virial 
overdensity leads to a universal halo mass function \citep{Despali2016}.

The number counts of convergence peaks with signal-to-noise $\nu(M,z)=\kappa_G/\sigma_{\rm noise}$ larger than a fixed 
threshold $\nu_{\rm th}$ is calculated as 
\begin{equation}
 N(>\nu_{\rm th})=\int {\rm d}z\,\frac{{\rm d}V}{{\rm d}z} \int {\rm d}M\,n(M) \mathcal{H}(\nu(M,z)-\nu_{\rm th})\,,
\end{equation}
in which ${\rm d}V/{\rm d}z$ is the appropriate comoving volume element per unit redshift and $\mathcal{H}(x)$ denotes 
the Heaviside step function. The rms variance of the noise field with a smoothing aperture $\theta_{G}$ is 
$\sigma_{\rm noise}=(\sigma_\epsilon^2/2)/2\pi\theta_{G}^2n_{\rm g}$ \citep{Kaiser1993,VanWaerbeke2000}, specified by 
the number density of source galaxies, $n_{\rm g}$, and the standard deviation of the intrinsic ellipticity 
distribution, $\sigma_\epsilon$. Finally, $\kappa_G$ is here averaged over the expected redshift distribution of 
sources, for which the parametrised form $p(z)\propto(z/z_0)^\alpha\exp[-(z/z_0)^\beta]$ from \cite{Smail1994} is used. 
For \textit{LSST} we adopted the values for the selection cut yielding a sample of galaxies with apparent limiting 
magnitude 
$i_{\rm AB}=26.8$, median redshift $z_{\rm g}=0.71$, and mean number density $n_{\rm g}=26$~arcmin$^{-2}$ 
\citep{Chang2013}, and considered as lenses the galaxy clusters in the redshift range $0.3<z<0.7$. 
For a \textit{Euclid}-like photometric survey with apparent limiting magnitude $H_{\rm AB}=24$ we use the same 
parameters as in \cite{Giannantonio2012} yielding a median redshift $z_{\rm g}=1$ and $n_{\rm g}=35$~arcmin$^{-2}$ 
\citep{Sartoris2016}, with galaxy clusters in the redshift range $0.3<z<1.0$. The modulation of number counts by the 
selection functions is not considered here, as going beyond the illustrative purposes of this application.

As shown in Table~\ref{tab:peakcounts}, in which the number counts of the convergence peaks per square degree with 
signal-to-noise ratio $3<\nu<5$ and $\nu>5$ are reported, the virialization model S$_{\rm T}$ is ruled out for every 
dark-energy model since it produces few peaks. Focusing on the standard virialization model S$_{\rm std}$ and peaks 
with $3<\nu<5$, the CPL, DE1, and DE2 dark-energy models produce $\sim10$ percent more ore less counts with respect to 
the $\Lambda$CDM; ODE and AS differ at 2-5~percent level; 2EXP and CNR yield sub-percent different counts; and these 
deviations increase by about 1.5 times for peaks with $\nu>5$. Interestingly enough, this trend between dark-energy 
models is not preserved when S$_1$ or S$_2$ virialization models are considered; here the number of peaks can differ 
up to about 65 percent with respect to the $\Lambda$CDM with standard virialization S$_{\rm std}$. It is worth to 
notice that the number counts per square degree with $\nu>5$ for the 2EXP and CNR models differ at the fourth 
decimal digit, so even wide-field surveys covering more than 10,000 square degrees cannot assess their difference 
because of Poisson noise. 
All these numbers actually strongly depend on the redshift distribution of background sources and on the redshift 
range of lenses, and neglect spatial clustering; they cannot therefore be used for accurate forecasts. 
An accurate study would finally also require to account for the degeneracy with the halo density profile and mass 
function's parameters.

\begin{table}
 \caption{Number counts of convergence peaks per square degree expected for \textit{LSST} and \textit{Euclid}-like 
weak-lensing 
 surveys with signal-to-noise ratio $3<\nu<5$ (upper table) and $\nu>5$ (lower table). The smoothing radius is 
 $\theta_{G}=2$ and 2.35~arcmin expected for \textit{LSST}- and \textit{Euclid}-like weak-lensing surveys (columns 2-5 
 and 6-9, respectively). 
 Spherical haloes have truncated NFW profile with the concentration parameter of \cite{Duffy2008} and are distributed 
 according to the FoF \cite{Watson2013} mass function.}
 \begin{center}
  \begin{tabular}{lcccccccc}
  \hline
          & \multicolumn{4}{c}{\textit{LSST}-like} & \multicolumn{4}{c}{\textit{Euclid}-like} \\
          & S$_{\rm std}$ & S$_1$ & S$_2$ & S$_{\rm T}$ & S$_{\rm std}$ & S$_1$ & S$_2$ & S$_{\rm T}$ \\
  \hline
  $\Lambda$CDM & 5.16 & 6.02 & 5.60 & 0.26 & 3.37 & 3.93 & 3.68 & 0.20 \\
  DE1          & 4.68 & 5.42 & 5.03 & 0.16 & 3.09 & 3.56 & 3.31 & 0.13 \\
  DE2          & 5.64 & 6.57 & 6.13 & 0.39 & 3.68 & 4.29 & 4.02 & 0.29 \\
  CPL          & 4.59 & 4.54 & 4.19 & 0.13 & 3.05 & 3.03 & 2.81 & 0.11 \\
  2EXP         & 5.16 & 6.02 & 5.60 & 0.27 & 3.37 & 3.93 & 3.68 & 0.21 \\
  AS           & 5.09 & 4.87 & 4.50 & 0.23 & 3.37 & 3.26 & 3.04 & 0.19 \\
  CNR          & 5.16 & 6.02 & 5.60 & 0.26 & 3.37 & 3.93 & 3.68 & 0.20 \\
  ODE          & 5.40 & 6.29 & 5.86 & 0.32 & 3.52 & 4.11 & 3.85 & 0.24 \\
  \hline
  $\Lambda$CDM & 0.214 & 0.304 & 0.263 & $<0.01$ & 0.156 & 0.219 & 0.192 & $<0.01$ \\
  DE1          & 0.183 & 0.254 & 0.218 & $<0.01$ & 0.134 & 0.185 & 0.161 & $<0.01$ \\
  DE2          & 0.249 & 0.355 & 0.308 & $<0.01$ & 0.180 & 0.254 & 0.223 & $<0.01$ \\
  CPL          & 0.193 & 0.193 & 0.164 & $<0.01$ & 0.144 & 0.146 & 0.126 & $<0.01$ \\
  2EXP         & 0.214 & 0.304 & 0.263 & $<0.01$ & 0.156 & 0.219 & 0.192 & $<0.01$ \\
  AS           & 0.229 & 0.218 & 0.187 & $<0.01$ & 0.169 & 0.165 & 0.143 & $<0.01$ \\
  CNR          & 0.214 & 0.304 & 0.263 & $<0.01$ & 0.156 & 0.219 & 0.192 & $<0.01$ \\
  ODE		   & 0.232 & 0.329 & 0.285 & $<0.01$ & 0.168 & 0.236 & 0.207 & $<0.01$ \\  
  \hline
  \end{tabular}
  \label{tab:peakcounts}
 \end{center}
\end{table}

\section{Conclusions}\label{sect:conclusions}

Starting from the basic equations for spherical collapse in an EdS universe as derived in \cite{Engineer2000} and 
refined in \cite{Shaw2008}, which naturally embed virialization by modelling shear and rotation of the (collisionless) 
matter overdensities, we generalised the formalism to arbitrary smooth dark-energy models.

Our aim was to study the features of the new approach and determine how it compares with the standard recipe. A 
standard result is, for example, that $\Delta_{\rm vir}$ is constant only for an EdS model and the value is usually 
used in determining haloes in $N$-body simulations. We wanted to answer the following questions: is $\Delta_{\rm vir}$ 
also constant in these recipes? How does the virial overdensity change? Is there a way to distinguish between the 
recipes? Note that it was not our intention to shed light on the nature of the dark-energy, since for realistic models 
we do not expect significant differences from $\Lambda$CDM. We wanted instead to understand, at a more fundamental 
level, how virialization works and this is only possible with a generic formalism that allows the study of various 
models.

In their original work \cite{Engineer2000} and \cite{Shaw2008} only outlined the methods but did not discuss how the 
virial overdensity evolves over time and how it compares with more standard approaches. We deemed important to repeat 
the calculation also for the EdS model as this allowed a more detailed investigation.

We studied the evolution of the radius of dark-matter overdensities reproducing the results of \cite{Engineer2000}, 
proving that for a $\Lambda$CDM model the turn-around occurs earlier since the initial overdensity is higher. 
Due to the stronger expansion caused by the cosmological constant term, the ratio $y_{\rm max}=r_{\rm ta}/r_{\rm vir}$ 
is larger for a $\Lambda$CDM model than for the EdS, as shown in Figure~\ref{fig:yq_EdS_LCDM}. Qualitatively the 
evolution of the normalised radius $y$ as a function of $q$ is independent of the background cosmology. 

Similar considerations arise for the peculiar velocity of the shells; see Figure~\ref{fig:hsc_EdS_LCDM}. 
In contrast with the radius, the maximum velocity is approximately the same for both the EdS and $\Lambda$CDM model. 
Very likely this conclusion is valid for more generic dark-energy models and justifies our choice of determining the 
value of $q$ by fitting the amplitude of $h_{\rm SC}$ to the value measured in EdS $N$-body simulations.

Our main result is the computation of the virial overdensity as a function of redshift for seven dark-energy models 
that match $\Lambda$CDM at redshift zero; see the list in Section~\ref{sect:DeltaV}. We showed that the formalisms of 
\cite{Engineer2000} and \cite{Shaw2008} can be easily extended to arbitrary smooth dark-energy models (see 
Appendix~\ref{sect:implementation} for the actual caveats about the numerical implementation). Qualitatively, the 
results are similar to those obtained by the standard virialization condition when the EdS model is considered, i.e., 
a non-varying value for $\Delta_{\rm vir}$. On the other hand for cosmologies with dark-energy $\Delta_{\rm vir}$ 
decreases with redshift until reaching lower values than EdS at late times, when dark-energy dominates; see 
Figure~\ref{fig:DeltaV}. 
The specific value depends in a non-trivial way on the dark-energy model and the virialization recipe. It is worth
noting that the virial overdensity for the Chevalier-Polarski-Linder dark-energy model is at odds with $\Lambda$CDM 
because of the rapid change of its equation-of-state at low redshift. Interestingly enough, for the recipe of 
\cite{Engineer2000}, the values of $\Delta_{\rm vir}$ at redshift $z=0$ are about 10 percent lower than when 
standard virialization is assumed, while for the recipe of \cite{Shaw2008} the difference is three times larger. 
Moreover, the range of values spanned by the different dark-energy models is reduced in a native model.

It is important to remember that all these virialization recipes are obtained from a statistical analysis of the 
peculiar velocity in $N$-body simulations, so mass information is lost. This implies that $\Delta_{\rm vir}$ is a 
function of time only, while in more realistic models one would also expect a mass-dependence. This is the case of the 
ellipsoidal model \citep{Angrick2010,Angrick2014}, in which the collapse is sensitive to the ellipticity and 
prolateness of the object.

Finally, it is also worth remarking that the actual value of the virial overdensity is pivotal for the definition of 
haloes in $N$-body simulations and observations that are based on the spherical overdensity algorithm. We argue that 
our results have an important impact in the determination of the halo mass function in the era of precision cosmology, 
in which several projects like \textit{Euclid} and \textit{LSST} are designed to assess any departure from the 
concordance 
$\Lambda$CDM model at an unprecedented level of accuracy.

As mainly driven by the dark matter component, weak gravitational lensing is very likely the ideal ground to assess 
the virialization mechanism and its degeneracy with dark-energy model. Being shear and rotation induced by the 
gravitational tidal field, an accurate description of the density profile of haloes and their outskirts and an 
accurate halo finder able to identify the dark matter haloes accounting for their triaxiality are necessary 
ingredients. 
Adopting well-established recipes, the radial surface-mass-density profile calculated for alternative dark-energy 
models shows up to $\sim 5-10$ percent difference with respect to $\Lambda$CDM near the core of haloes for clusters at 
redshift $0.3-1$ regardless of their mass and concentration, with smaller dependence on the virialization model. 
High-resolution deep imaging surveys would reveal the differences exhibited in these models. 
Larger deviations are expected by the counts of convergence peaks, whose usefulness to investigate dark-energy models 
with standard virialization was already established by \cite{Weinberg2003}. Here we proved that the differences of 
peaks' counts among several, non-trivial dark-energy models are fully degenerate with the specific model of 
virialization. 
Despite the many approximations we adopted, this is a further proof that precision cosmology for the dark sector 
requires an accurate description of gravitational clustering on megaparsec scales. Wide-field photometric surveys such 
as \textit{LSST} and \textit{Euclid} will be excellent of revealing the subtle differences expected by less trivial 
dark-energy 
models.

Finally, we remind the reader that the other critical quantity in the spherical collapse model is the linear 
overdensity at collapse, $\delta_{\rm c}$, determined by the initial conditions such that $\delta\rightarrow\infty$ at 
the chosen collapse time. As explained in Appendix~\ref{sect:implementation}, for the calculation we adopt here the 
standard non-linear equation, the evolution of $\delta_{\rm c}$ is therefore unchanged and does not require further 
exploration; see \cite{Pace2010,Pace2012} for the details.

The numerical code is available upon request.

\appendix

\section{Ordinary SCM and virialization in Einstein-de Sitter and \texorpdfstring{$\boldsymbol{\Lambda}$CDM}{LCDM} 
models}\label{appendix:SCM}

We remind here the results for EdS and $\Lambda$CDM models of the ordinary SCM. Using the dimensionless scale factor 
$x=a/a_{\rm ta}$ and radius $y=R/R_{\rm ta}$ rescaled to their values at the turn-around, and the dimensionless time 
parameter $\tau\equiv H_{\rm ta}t$, the first Friedmann equation and Eq.~(\ref{eqn:Rddot}) can be written as
\begin{align}
 \frac{\mathrm{d}x}{\mathrm{d}\tau} = &\, \left[\frac{\omega}{x} + \lambda x^2g(x) + (1-\omega-\lambda)\right]^{1/2}\,, 
 \label{eqn:xdot}\\
 \frac{\mathrm{d}^2y}{\mathrm{d}\tau^2} = &\, -\frac{\omega\zeta_{\rm ta}}{2y^2} - 
 \frac{1+3w_{\rm de}(x)}{2}\lambda g(x)y\,,\label{eqn:yddot}
\end{align}
where $\omega$ and $\lambda$ denote respectively the matter and dark-energy parameters at turn-around, $g(x)$ the 
evolution of the dark-energy density relative to turn-around with equation-of-state parameter $w_{\rm de}$,
\begin{equation}
 g(x) = \exp{\left\{-3\int_1^x[1+w_{\rm de}(x^{\prime})]~{\rm d}\ln{x^{\prime}}\right\}}\,,
\end{equation}
and $\zeta_{\rm ta}$ is the non-linear overdensity of the collapsing sphere with respect to the background at the 
turn-around (for an EdS model $\zeta_{\rm ta}=\left({3\pi}/{4}\right)^2\simeq 5.55$; \cite{Kihara1968}). Remark that 
without shear and rotation, $R$ is the actual radius of spherical perturbations.

Analytical solutions of equations (\ref{eqn:xdot}-\ref{eqn:yddot}) can be obtained for spatially flat models with 
constant $w_{\rm de}$ \citep{Lee2010,Lee2010c,Pace2017a}. The essence of the virialization process in the ordinary SCM 
is already captured by the EdS model; according to the virial theorem, the potential energy 
$U=-\tfrac{3}{5}\frac{GM}{R}$ yields $y_{\rm vir}\equiv R_{\rm vir}/R_{\rm ta}=1/2$, which ultimately gives
\begin{equation*}
 \Delta_{\rm vir}(a_{\rm c}) = 18\pi^2 \simeq 177.65\,, \quad 
 \Delta_{\rm vir}(a_{\rm vir}) = 18\pi^2\left(\frac{3}{4}+\frac{1}{2\pi}\right)^2 \simeq 146.84\,,
\end{equation*}
for the values of overdensities at collapse and virialization times, respectively. 
The corresponding linear extrapolated overdensities are the well-known values
\begin{equation*}
 \delta_{\rm c}(a_{\rm c}) = \frac{3}{20}(12\pi)^{2/3} \simeq 1.686 \,, \quad 
 \delta_{\rm c}(a_{\rm vir}) = \frac{3}{20}(6+9\pi)^{2/3} \simeq 1.58 \,.
\end{equation*}

In a $\Lambda$CDM scenario the picture is more complicated because of the potential energy of the cosmological 
constant, which leads to a third order algebraic equation in the rescaled radius $y$. Following 
\cite{Wang1998,Maor2005},\footnote{As also noted by \cite{Meyer2012}, different strategies have also been adopted 
leading to the same results, at least for a $\Lambda$CDM model; see however \cite{Wang2006}.} supposing that the 
solution is only slightly different from the EdS, the virial condition 
$\left[U + \frac{R}{2}\frac{\partial U}{\partial R}\right]_{\rm vir} = U_{\rm ta}$, can be solved perturbatively, 
yielding at first order
\begin{equation}
 y_{\rm vir} = \frac{1-\eta_{\rm vir}/2}{2+\eta_{\rm t}-3\eta_{\rm vir}/2}\,,
\end{equation}
where
\begin{equation*}
 \eta_{\rm t} = 2\zeta_{\rm ta}^{-1}\frac{\Omega_{\Lambda}(a_{\rm ta})}{\Omega_{\rm m}(a_{\rm ta})}\,, \quad 
 \eta_{\rm vir} = 2\zeta_{\rm ta}^{-1}\left(\frac{a_{\rm ta}}{a_{\rm vir}}\right)^3
                \frac{\Omega_{\Lambda}(a_{\rm vir})}{\Omega_{\rm m}(a_{\rm vir})}\,.
\end{equation*}
Finally, the virial overdensity with respect to the background density is defined as in \citep{Wang1998},
\begin{equation}
 \Delta_{\rm vir} = \zeta_{\rm ta}\left(\frac{x_{\rm c,v}}{y_{\rm vir}}\right)^3\,,
\end{equation}
where $\zeta_{\rm ta}$ has been defined above, and $x_{\rm c}$  and $x_{\rm vir}$ are respectively the scale factors 
at collapse and virialization normalised at the turn-around epoch.
Note that the crucial quantity is $\eta_{\rm vir}$, which is evaluated at the virialization time $a_{\rm vir}$. 
The latter should be determined iteratively \citep{Meyer2012,Pace2017a}, but it is routinely approximated by the 
collapse time, $a_{\rm vir}\simeq a_{\rm c}$. However, regardless the exact epoch considered, only the value of 
$\Delta_{\rm vir}$ will change but not its qualitative evolution \citep{Pace2017a}.

\section{Extended SCDM -- algorithm and virialization in smooth dark-energy models}\label{sect:implementation}

The general architecture of our code, publicly available upon request, is based on the one described in Appendix~A of 
\cite{Pace2017a}. 
Note that we detail the initial conditions used to solve the equations of the SCM, not clearly stated in 
\cite{Engineer2000}.\footnote{For this reason we have been able to only approximately infer that the parametrisation 
for $S(\delta)$ by \cite{Engineer2000} should be valid for $\delta\gtrsim 15$, which explains why we do not reproduce 
exactly their values.}

\begin{enumerate}
\item \textit{Collapse epoch and initial conditions.} For any cosmological model defined by $\Omega_{\rm m,0}$, 
      $\Omega_{\rm de,0}$, and $w_{\rm de}(z)$, the collapse redshift $z_{\rm c}$ is determined by solving the 
      standard equations of the SCM as explained in \cite{Pace2017a}. This value fixes the initial values 
      $(\delta_{\rm in}, \delta_{\rm in}^{\prime})$ and determines the values of $\zeta_{\rm ta}$, $z_{\rm vir}$, 
      $y_{\rm vir}$, $x_{\rm vir}$, and finally $\Delta_{\rm vir}$ for the ordinary SCM.
\item \textit{Effective radius of the sphere $y$: solve Eqs.~(\ref{eqn:yq})}. Differently from the standard SCM, 
      the effective radius does not vanish at virialization but tends to the virial radius. Eqs.~(\ref{eqn:yq}) 
      can therefore be solved with final conditions $y(z_{\rm c})=1$, $y^{\prime}(z_{\rm c})=0$, going 
      backward-in-time from the collapse till the time used to integrate forwards the equations describing the 
      evolution of $\delta$.
\item \textit{Approach I: Evolution of the perturbation overdensity $\delta$ with virialization terms S$_1$, S$_2$ -- 
      Equations~(\ref{eqn:S12}).} The value of $q$ is determined by matching the amplitude of the theoretical 
      prediction of $h_{\rm SC}$ to the fitting function of \cite{Hamilton1991} (see Table~\ref{tab:qpar}), getting 
      the corresponding value of $\delta_{h}$ where the matching is realised. The equations of 
      motion~(\ref{eqn:deltaS}) are then integrated using the initial conditions for $\delta$ obtained in step (i), 
      using the equations of the SCM with $S(\delta)=0$ for $\delta<\delta_{h}$, and the shear-rotation virialization 
      term S$_1(\delta)$ or S$_2(\delta)$ for $\delta>\delta_{h}$. 
      The integration proceeds till the collapse time $z_{\rm c}$.
\item \textit{Approach II: Evolution of the perturbation overdensity $\delta$ with virialization term S$_{\rm T}$ -- 
      Eq.~(\ref{eqn:ST})}. Since an analytical relation $\eta=\eta(\delta)$ does not exist, Eq.~(\ref{eqn:ST}) 
      cannot be used directly and a parametric relation is needed. Furthermore, the function $T(\tau)$ is 
      complex-valued for very large values of $\eta$. A safe numerical implementation proceeds as follows:
      \begin{enumerate}
       \item find the value $\eta_f$ such that $\tau(\eta_f)\equiv\tau_f=5.516$;
       \item sample the functions $\tau(\eta)$ and correspondingly $1+\delta(\eta)$ defined in 
             Equation~(\ref{eqn:delta_T}) in the range $10^{-3}\leqslant\eta\leqslant\eta_f-10^{-3}$;
       \item solve the full Eq.~(\ref{eqn:nldeltaa}) for $\delta$ including the virialization term from 
             Eq.~(\ref{eqn:ST}): for each value of $a$ and $\delta$, evaluate the function $(1+\delta)/f(a)$, 
             determine the corresponding values of $\eta$ and $\tau$, finally use them in Eq.~(\ref{eqn:ST}) to 
             determine $S(\delta)$. This procedure is repeated at each time step till $z_{\rm c}$. Note that for some 
             models the default sampling of the tabulated function is insufficient to return a correct value. In these 
             cases, we set $\eta$ to its maximum tabulated value.
      \end{enumerate}
\item Indicating with $\delta_{\rm NL}$ the value of $\delta$ at $z_{\rm c}$, the virial overdensity is 
      $\Delta_{\rm vir}=1+\delta_{\rm NL}$.
\end{enumerate}

\section*{Acknowledgements}
\noindent FP acknowledges support from the STFC grant ST/P000649/1. 
DFM thanks the Research Council of Norway for their support, and the resources provided by UNINETT Sigma2 -- the 
National Infrastructure for High Performance Computing and Data Storage in Norway. 
CS thanks Nicolas Martinet for useful discussions. 
This paper is based upon work from the COST action CA15117 (CANTATA), supported by COST (European Cooperation in 
Science and Technology). The authors thank Damien Trinh and Matthew Pieri for carefully reading the paper and 
improving its style. FP would like to thank Bj\"orn Sch\"afer for useful discussions on the Birkhoff's theorem. The 
authors would also like to thank an anonymous referee for carefully reading this manuscript and for his/her comments 
which helped to improve the scientific quality of this work.

\bibliographystyle{JHEP}
\bibliography{spcVirialization.bbl}

\label{lastpage}

\end{document}